\documentclass[10pt]{article}
\usepackage{amsmath,amssymb,amsthm,bm,graphicx,natbib}

\long\def\comment#1{}

\title{Bayesian ICA-based source separation of Cosmic Microwave Background by a discrete functional approximation}
\author{Simon P.~Wilson\thanks{School of Computer Science and Statistics, Trinity College, Dublin 2, Ireland. \texttt{swilson@tcd.ie}} \and Ji Won~Yoon \thanks{School of Computer Science and Statistics, Trinity College, Dublin 2, Ireland. \texttt{yoonj@tcd.ie}}}
\date{}

\begin{document}

\maketitle

\newcommand{\fix}{\marginpar{FIX}}
\newcommand{\new}{\marginpar{NEW}}

\begin{abstract}
A functional approximation to implement Bayesian source separation analysis is introduced and applied to separation of the Cosmic Microwave Background (CMB) using WMAP data. The approximation allows for tractable full-sky map reconstructions at the scale of both WMAP and Planck data and models the spatial smoothness of sources through a Gaussian Markov random field prior.  It is orders of magnitude faster than the usual MCMC approaches.  The performance and limitations of the approximation are also discussed.

\end{abstract}

\section{Introduction}

Source separation is one of the initial data processing tasks for multi-channel image data, such as have been obtained at microwave frequencies by COBE, WMAP and more recently Planck. The goal in this case, and the application of focus for this paper, is to reconstruct the CMB signal by separating it from other sources.  Additionally, maps of the other sources may be obtained and of interest.

In this paper we propose a new method of implementing Bayesian inference to source separation, based on a discrete grid approximation to a posterior density,  and apply it to CMB data. The method is substantially faster than the usual sampling-based approaches to Bayesian inference, allows for full-sky source reconstructions of data of the size of WMAP ($\approx 3\times10^6$ pixels at 5 channels) in practical amounts of time, and should remain feasible for data that is an order of magnitude larger, at the higher resolution of Planck. Further, the approach permits spatially smooth priors to be specified for the sources through a Gaussian Markov random field.

Bayesian source separation computes the posterior distribution of source components given data. Several approaches based on factor analysis or independent components analysis (ICA) models have been proposed e.g.\ \cite{hobson98,eriksen06,wilson08a,kuruoglu10}. The advantages of the Bayesian approach are the ease with which domain knowledge can be used in the analysis through the specification of the prior distribution, and the coherent treatment of uncertainty which leads to proper estimation of the uncertainties in the source components from the uncertainties in the model and data. The former is particularly useful in this context as so much is known about the sources and how they contribute to the data maps at different channels; the incorporation of such information can greatly improve the separation \citep{wilson08a}.

The principal disadvantage of Bayesian methods is computational.  The usual approach to computing the posterior distribution is through Markov chain Monte Carlo (MCMC) sampling \citep[Chapter 11]{gelman03}. For the model considered here, the computational and storage requirements of an MCMC solution make it impractical to consider separation at the scale of complete maps of WMAP data, as well as implementation of standard statistical diagnostics for model assessment like cross validation, even with partitioning the data into smaller regions.   MCMC can also suffer from problems of slow convergence and exploration of the support of the posterior distribution. Nevertheless there has been some progress in MCMC methods; \cite{kayabol09a} implemented a Metropolis algorithm for a non-Gaussian Markov random field prior on the sources, which in \cite{kayabol09b} is speeded up considerably by the use of a Langevin sampler.

Functional approximations to high-dimensional posterior distributions, rather than sampling-based approximations, are an alternative that have gained some prominence in the last 10 years or so.  They can be substantially faster than MCMC. Variational Bayes (VB) is one approximation that has seen some application to source separation \citep{winther07,cemgil07}, where the idea is to find an approximating distribution to the posterior that is close in the sense of Kullback-Leibler divergence \citep{jordan98}. VB relies on factorising the approximating distribution for a tractable algorithm, which tends to lead to under-estimation of posterior variances, although means are in general approximated well \citep{wang04}. 

This paper demonstrates that in fact a relatively unsophisticated discrete approximation is sufficient in the case of a Gaussian likelihood and a Gaussian Markov random field prior for the sources, as long as the number of hyperparameters in the model is not too large.  Later we discuss how these assumptions can be relaxed by generalising the approximation to the integrated nested Laplace approximation (INLA) of \cite{rue08}. Although our approach is restricted to a much smaller class of models than VB, INLA has been shown to be both fast and accurate within this class.  

Section \ref{sec:model} describes the factor analysis model and Section \ref{sec:prior} discusses prior specification for the Bayesian inference. Section \ref{sec:postcalc} describes the approximation that allows computation of an approximation to the posterior mean of the sources, which is then illustrated in Section \ref{sec:wmap} by analysis of the 7-year WMAP data into 4 sources.


\section{Model}
\label{sec:model}

The data consist of images of intensities at $n_{f}$ frequencies $v_{1}, \cdots, v_{n_{f}}$ over the sky at $J$ pixels.  The data at pixel $j$ are denoted $\bm{y}_{j}\in \mathbb{R}^{n_{f}}$, $j=1, 2, \cdots, J$, while $\bm{Y}_k = ( y_{1k},\ldots,y_{Jk})^T$ denotes the all-sky image at frequency $\nu_k$.  There are $n_s$ sources. The vector of source components at pixel $j$  is denoted $\bm{s}_{j}\in \mathbb{R}^{n_{s}}$ and the image of source $i$ is $\bm{S}_i = (s_{1i},\ldots,s_{Ji})^T$.

We assume the standard statistical independent components analysis model for $\bm{y}_j$:
\begin{equation}
\bm{y}_{j} = \bm{A \, s}_j + \bm{e}_j, \; j=1,\ldots,J,
\label{eq:fa_pixel}
\end{equation}
where $\bm{A}$ is an $n_{f}\times n_{s}$ mixing matrix and $\bm{e}_{j}$ is a vector of $n_{f}$ independent Gaussian error terms with precisions $\bm{\tau} =(\tau_{1}, \cdots, \tau_{n_{f}})$.

Stacking the $\bm{Y}_k$ and $\bm{S}_i$ as $\bm{Y} = (\bm{Y}_1^T,\ldots,\bm{Y}_{n_f}^T)^T$ and $\bm{S} = (\bm{S}_1^T,\ldots,\bm{S}_{n_s}^T)^T$, and stacking the error terms by frequency $\bm{E}=(e_{11},\ldots,e_{J1},e_{12},\ldots,e_{Jn_f})$, Eq.\ \ref{eq:fa_pixel} can be rewritten:
\begin{equation}
\bm{Y} = \bm{B} \, \bm{S} + \bm{E},
\label{eq:fa_stacked}
\end{equation}
where $\bm{B} =   \bm{A} \otimes \bm{I}_{J\times J}$ is the Kronecker product of $\bm{A}$ with the $J\times J$ identity matrix.  

It is common for each pixel to be observed more than once, and the scanning schedule of the detector means that different pixels may be observed a different number of times. Define $n_j$ to be the number of times that pixel $j$ is observed. Where this occurs, the Gaussian error assumption implies that the probability distribution for the $n_j$ observations of $y_{jk}$ is equivalent to a single observation that is the average of the observations with precision $e_{jk} = n_j \tau_k$; like this we consider each $y_{jk}$ to be observed once and $\bm{E}$ is zero-mean Gaussian with precision matrix 
\begin{equation*}
\bm{C} = \mbox{diag}(n_1\tau_1,\ldots,n_J\tau_1,\ldots,n_1 \tau_{n_f},\ldots,n_J \tau_{n_f}).
\end{equation*}  
Uniqueness of the solution for $\bm{A}$ and $\bm{S}$ is forced by setting a row of $\bm{A}$ (the fourth row here) to be ones. 

Four sources are assumed in this work: CMB, synchrotron, galactic dust and free-free emission.  A parameterisation of $\bm{A}$ is assumed, following \cite{eriksen06}. The first column of $\bm{A}$ is the contribution of CMB and is assumed known (black body):
\[ A_{k1} = g(\nu_k)/g(\nu_4), \: k=1,\ldots,n_f, \mbox{ where } g(\nu_k) = \left(\frac{\eta \nu_k}{k_{B}T_{0}}\right)^{2} \frac{\exp({\eta \nu_k/k_{B}T_{0}})}{(\exp(\eta \nu_k/k_{B}T_{0})-1)^{2}}, \]
$T_{0} = 2.725K$ is the average CMB temperature, $\eta$ is the Planck constant and $k_{B}$ is Boltzmann's constant. It has no free parameter. 
The second column is for synchrotron radiation and has entries of the form
\[ A_{k2} = \left(\frac{\nu_{k}}{\nu_{4}}\right)^{\theta_{s}} \]
for a free parameter $\theta_s$, the third column is for galactic dust and has entries of the form
\[ A_{k3} =  \frac{\exp(\eta \nu_{4}/k_{B}T_{1})-1}{\exp(\eta \nu_{k}/k_{B}T_{1})-1}\left(\frac{\nu_{k}}{\nu_{4}}\right)^{1+\theta_{d}} \]
for a free parameter $\theta_d$, where $T_1 = 18.1 K$, and the fourth column is for free-free emission and has entries of the form
\[ A_{k4} = \left(\frac{\nu_{k}}{\nu_{4}}\right)^{-2.19} \]
and has no free parameter.  Hence $\bm{A}$ is parameterised by $\theta_s$ and $\theta_d$.


\section{Prior}
\label{sec:prior}

\paragraph{Prior for $\bm{\theta_s,\theta_d}$:} Prior studies give ranges for the mixing matrix parameters: $-3.0 < \theta_s < -2.3$ and $1.0 < \theta_d < 2.0$ \citep{eriksen06}. This information is quantified as independent uniform distributions on these ranges.

\paragraph{Prior for the sources:} Independent intrinsic Gaussian Markov random field (GMRF) priors are used for each source $\bm{S}_i$.  These priors impose spatial smoothness on $\bm{S}_i$ by inducing conditional independence of a pixel on the others given its neighbours. In this paper the first order intrinsic GMRF \citep[Chapter 3]{rue05} is used, which imposes that the differences
\[ \sum_{j^{\prime} \in c(j)} (S_{ij} - S_{ij^{\prime}}) \]
are independent zero-mean Gaussian with precision $\phi_i$, where $c(j)$ is the set of pixel indices of the four nearest neighbours of pixel $j$. This leads to a distribution of $\bm{S}_i$ that is of zero-mean multivariate Gaussian form:
\begin{equation}
p(\bm{S}_i \, | \, \phi_i) \: \propto \:  |\bm{Q}(\phi_i)|^{0.5} \: \exp\left(-0.5\bm{S}_i^T \, \bm{Q}(\phi_i) \, \bm{S}_i \right),
\label{eq:igmrf}
\end{equation}
where $\bm{Q}(\phi_i)$ is a $J \times J$ matrix that can be written as $\bm{Q}(\phi_i) = \phi_i \bm{D}^T \bm{D}$, where the elements of $\bm{D}$ are defined as:
\begin{equation*}
D_{j_1,j_2} = \begin{cases} 1, & \mbox{if } j_2 \in c(j_1) \\ 0, & \mbox{otherwise,} \end{cases}
\end{equation*}
for $j_1 \neq j_2$ and main diagonal elements are $D_{jj} = -\sum_{\substack{l=1\\l \neq j}}^J D_{j,l}$. The term intrinsic GMRF comes from the fact that $\bm{Q}(\phi_i)$ is not of full rank, hence Equation \ref{eq:igmrf} is not a well-defined probability density function. However, the posterior distributions of the $\bm{S}_i$ will still be properly defined; again, see Chapter 3 of \cite{rue05}.  

Let $\bm{\Psi} = (\theta_d,\theta_s,\phi_1,\ldots,\phi_{n_s})$ denote all the hyperparameters in the model.  The distribution of the stacked vector of sources is then
\begin{equation}
p(\bm{S} \, | \, \bm{\Psi}) \: \propto \: |\bm{Q}(\bm{\Psi})|^{0.5} \: \exp\left(-0.5\bm{S}^T \, \bm{Q}(\bm{\Psi})\, \bm{S} \right),
\label{eq:S_prior}
\end{equation} 
where 
\[ \bm{Q}(\bm{\Psi}) = \left( \begin{array}{cccc} \bm{Q}(\phi_1) & \bm{0} & \cdots & \bm{0} \\ 0 & \bm{Q}(\phi_2) & \cdots & \bm{0} \\ \vdots & \vdots & \ddots & \vdots \\ \bm{0} & \bm{0} & \cdots & \bm{Q}(\phi_{n_s}) \end{array} \right). \]

\paragraph{Prior for the $\bm{\phi_i}$:} Independent gamma distributions are used. The density function is of the form $p(\phi_i) \propto \phi_i^{b_i-1} e^{-a_i \phi_i}$ for positive hyperparameters $a_i$ and $b_i$. Default non-informative values are $b_i=1$ and $a_i$ very small, otherwise prior knowledge about the degree of variation in each source can inform the choice using, for example, that the mean and standard deviation of this distribution are $b_i/a_i$ and $\sqrt{b_i}/a_i$ respectively.

\paragraph{Prior for the $\bm{\tau_k}$:} The $\tau_k$ are assumed known. This is a reasonable assumption for microwave maps, based on data from detector calibration.


\section{Posterior Calculations}
\label{sec:postcalc}
The unknown quantities in this model are $\bm{S}$ and $\bm{\Psi}$. The posterior distribution is then:
\begin{eqnarray} p(\bm{S},\bm{\Psi} \, | \, \bm{Y}) & \propto & p(\bm{Y} \, | \, \bm{S},\bm{\Psi}) \, p(\bm{S} \, | \, \bm{\Psi}) \, p(\bm{\Psi}) \nonumber \\
& = & p(\bm{Y} \, | \, \bm{S},\theta_d,\theta_s) \, \left( \prod_{i=1}^{n_s} p(\bm{S}_i \, | \, \phi_i)\right) \: \left( p(\theta_s) \, p(\theta_d) \, \prod_{i=1}^{n_s} \, p(\phi_i) \right); \label{eq:posterior}
\end{eqnarray}
all these terms are defined in Sections  \ref{sec:model} and \ref{sec:prior}.

The aim is to compute $\mathbb{E}(\bm{S} \, | \, \bm{Y})$, the posterior expectation of the sources.  For this, an approximation to the marginal posterior distribution of $\bm{\Psi}$ is needed first.

\subsection{Discrete approximation of $\bm{p(\Psi \, | \, Y)}$}
\label{subsec:inla}
Simple manipulation of the multiplicative law of probability shows that for any $\bm{S}$ such that $p(\bm{S} \, | \, \bm{Y}, \bm{\Psi}) > 0$,
\begin{equation} 
p(\bm{\Psi} \, | \, \bm{Y}) \: \propto \:  \frac{p(\bm{Y} \, | \, \bm{S},\bm{\Psi}) \, p(\bm{S} \, | \, \bm{\Psi}) \, p(\bm{\Psi})}{p(\bm{S} \, | \, \bm{Y}, \bm{\Psi})}.
\label{eq:inla_basicapprox} 
\end{equation}
The numerator terms of the right side of Eq.\ \ref{eq:inla_basicapprox} are given in Eq.\ \ref{eq:posterior} and the denominator term is easily shown to be Gaussian:
\begin{equation}
p(\bm{S} \, | \, \bm{Y}, \bm{\Psi})  \: = \: (2\pi)^{0.5n_s J} \, |\bm{Q}^*(\bm{\Psi})|^{0.5} \: \exp(-0.5(\bm{S}-\bm{\mu}^*(\bm{\Psi}) )^T \bm{Q}^*(\bm{\Psi}) (\bm{S}-\bm{\mu}^*(\bm{\Psi}) )),
\label{eq:inla_denom}
\end{equation}
where the precision and mean are
\begin{eqnarray}
\bm{Q}^*(\bm{\Psi}) &=& \bm{Q}(\bm{\Psi}) + {\bf B}^{T}{\bf C}{\bf B} \mbox{ and} \label{eq:Q*} \\
\bm{\mu}^{*}(\bm{\Psi})  &=& \bm{Q}^*(\bm{\Psi})^{-1}{\bf B}^{T}{\bf C}{\bf Y} \label{eq:mu*}
\end{eqnarray}
respectively. A numerically stable value of $\bm{S}$ at which to evaluate Eq.\ \ref{eq:inla_basicapprox} is $\arg \max_{\bm{S}} p(\bm{S} \, | \, \bm{Y}, \bm{\Psi}) = \bm{\mu}^*(\bm{\Psi})$, which gives the definition of a function $q(\bm{\Psi} \, | \, \bm{Y})$:
\begin{equation*}
p(\bm{\Psi} \, | \, \bm{Y}) \: \propto \:  |\bm{Q}^*(\bm{\Psi})|^{-0.5} \: p(\bm{Y} \, | \, \bm{S}=\bm{\mu}^*(\bm{\Psi}) ,\bm{\Psi}) \, p(\bm{S}=\bm{\mu}^*(\bm{\Psi})  \, | \, \bm{\Psi}) \, p(\bm{\Psi}) =  q(\bm{\Psi} \, | \, \bm{Y}).
\label{eq:inla_q}
\end{equation*}
Evaluation of $q(\bm{\Psi} \, | \, \bm{Y})$ requires the computation of the determinant and inverse of $\bm{Q}^*(\bm{\Psi})$ whose dimension ($n_sJ \times n_sJ$) is prohibitively large.   What is possible is to define $q_W$ over a smaller window $W$ of pixels:
\begin{equation*}
q_W(\bm{\Psi} \, | \, \bm{Y}_W) \: = \: | \bm{Q}^*_W(\bm{\Psi})|^{-0.5} \: p(\bm{Y}_W \, | \, \bm{S}_W=\bm{\mu}^*_W(\bm{\Psi}) ,\bm{\Psi}) \, p(\bm{S}_W=\bm{\mu}^*_W(\bm{\Psi})  \, | \, \bm{\Psi}) \, p(\bm{\Psi}),
\end{equation*}
where $\bm{Y}_W$ and $\bm{S}_W$ are the elements of $\bm{Y}$ and $\bm{S}$ over the pixels in $W$. The matrix $\bm{Q}^*_W(\bm{\Psi})$ is the precision matrix of $\bm{S}_W$ given $\bm{Y}_W$ and $\bm{\Psi}$ and follows Eq.\ \ref{eq:Q*} with $\bm{Q}(\bm{\Psi})$, $\bm{B}$ and $\bm{C}$ replaced by their submatrices $\bm{Q}_W(\bm{\Psi})$, $\bm{B}_W$ and $\bm{C}_W$ corresponding to the pixels in $W$;  $\bm{\mu}^*_W(\bm{\Psi}) = \bm{Q}^*_W(\bm{\Psi})^{-1}{\bf B}_W^{T}{\bf C}_W{\bf Y}_W$ follows Eq.\ \ref{eq:mu*} similarly.  The size of the window $W$ is chosen so that both $|\bm{Q}^*_W(\bm{\Psi})|$ and $\bm{Q}^*_W(\bm{\Psi})^{-1}$ can be computed.

Now $p(\bm{\Psi} \, | \, \bm{Y}_W)$ can be derived numerically by computing the proportionality constant 
\[ \left( \int_{\forall \Psi}  q_W(\bm{\Psi} \, | \, \bm{Y}_W) \: d\bm{\Psi} \right)^{-1}\]
to obtain it from  $q_W$. This is done by evaluating $q_W$ over a discrete set $\cal{Q}$ of values of $\bm{\Psi}$. The set is defined by an initial exploration of $q_W$ to find a mode with respect to $\bm{\Psi}$, then exploring around that mode to find a high probability region. Here, the Hessian of $\log(q_W(\bm{\Psi} \, | \, \bm{Y}))$ with respect to $\bm{\Psi}$ is computed at the mode and $\cal{Q}$ formed by taking points out along each parameter axis, at intervals equal to the square root of the inverse of the Hessian, until $\log(q_W(\bm{\Psi} \, | \, \bm{Y}))$ is 3 less than its value at the mode.   \cite{rue08} recommend exploring along the eigenvectors of the Hessian, which may be more efficient. The proportionality constant is approximated by the Riemann sum over $\cal{Q}$ thus:
\begin{equation}
p(\bm{\Psi} \, | \, \bm{Y}_W) \: \approx \: \frac{q_W(\bm{\Psi} \, | \, \bm{Y}_W)}{\sum_{\bm{\Psi} \in \cal{Q}} q_W(\bm{\Psi} \, | \, \bm{Y}_W) \: \Delta\bm{\Psi}}, \; \bm{\Psi} \in \cal{Q},
\label{eq:inla_p_constant}
\end{equation}
where $\Delta \bm{\Psi}$ are volume weights. 

\subsection{Approximate evaluation of $\bm{\mathbb{E}(S \, | \, Y)}$}
The sources are reconstructed using the posterior means. In principal one wants to compute the $\mathbb{E}(S_{ij} \, | \, \bm{Y})$ but this would require the evaluation of $q(\bm{\Psi} \, | \, \bm{Y})$. Instead, the posterior means over $W$ are computed via the conditional expectation formula and Eq.\ \ref{eq:inla_p_constant}:
\begin{eqnarray}
\mathbb{E}(\bm{S}_{W} \, | \, \bm{Y}_W) & = & \mathbb{E}_{\bm{\Psi}|\bm{Y}_W} ( \mathbb{E}(\bm{S}_{W} \, | \, \bm{Y}_W,\bm{\Psi})) \nonumber \\
& = & \int_{\bm{\Psi}} \bm{\mu}^*_{W}(\bm{\Psi}) \: p(\bm{\Psi} \, | \, \bm{Y}_W) \: d\bm{\Psi} \nonumber \\
& \approx & \frac{\sum_{\bm{\Psi} \in \cal{Q}} \bm{\mu}^*_{W}(\bm{\Psi}) \: q_W(\bm{\Psi} \, | \, \bm{Y}_W) \: \Delta\bm{\Psi}}{\sum_{\bm{\Psi} \in \cal{Q}}  \: q_W(\bm{\Psi} \, | \, \bm{Y}_W) \: \Delta\bm{\Psi}},
\label{eq:inla_postmean}
\end{eqnarray}
and these used as an approximation to $\mathbb{E}(S_{ij} \, | \, \bm{Y})$ for any $j \in W$.

\section{Simulated data example}
\label{sec:simdata}
\begin{figure}
\centering
\includegraphics[height = 40mm, width=40mm]{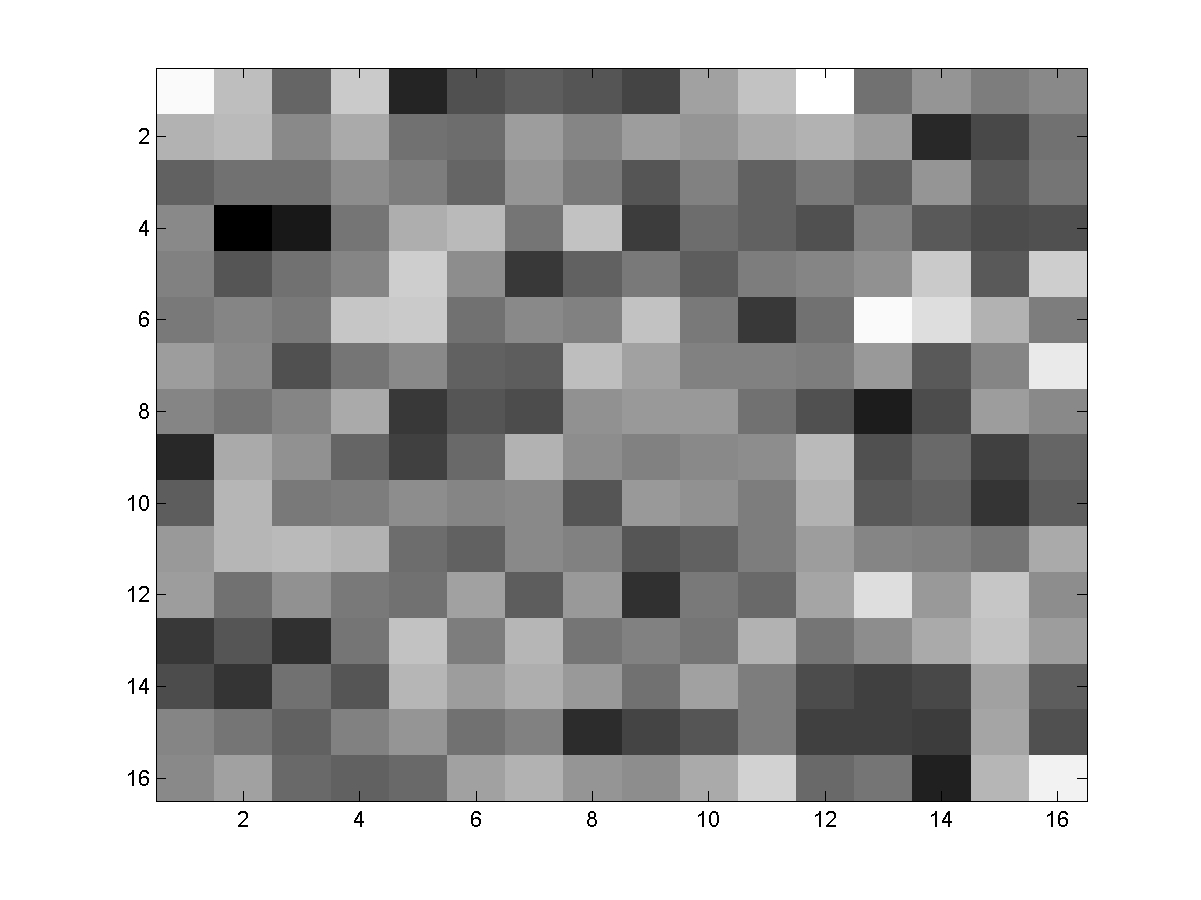}
\includegraphics[height = 40mm, width=40mm]{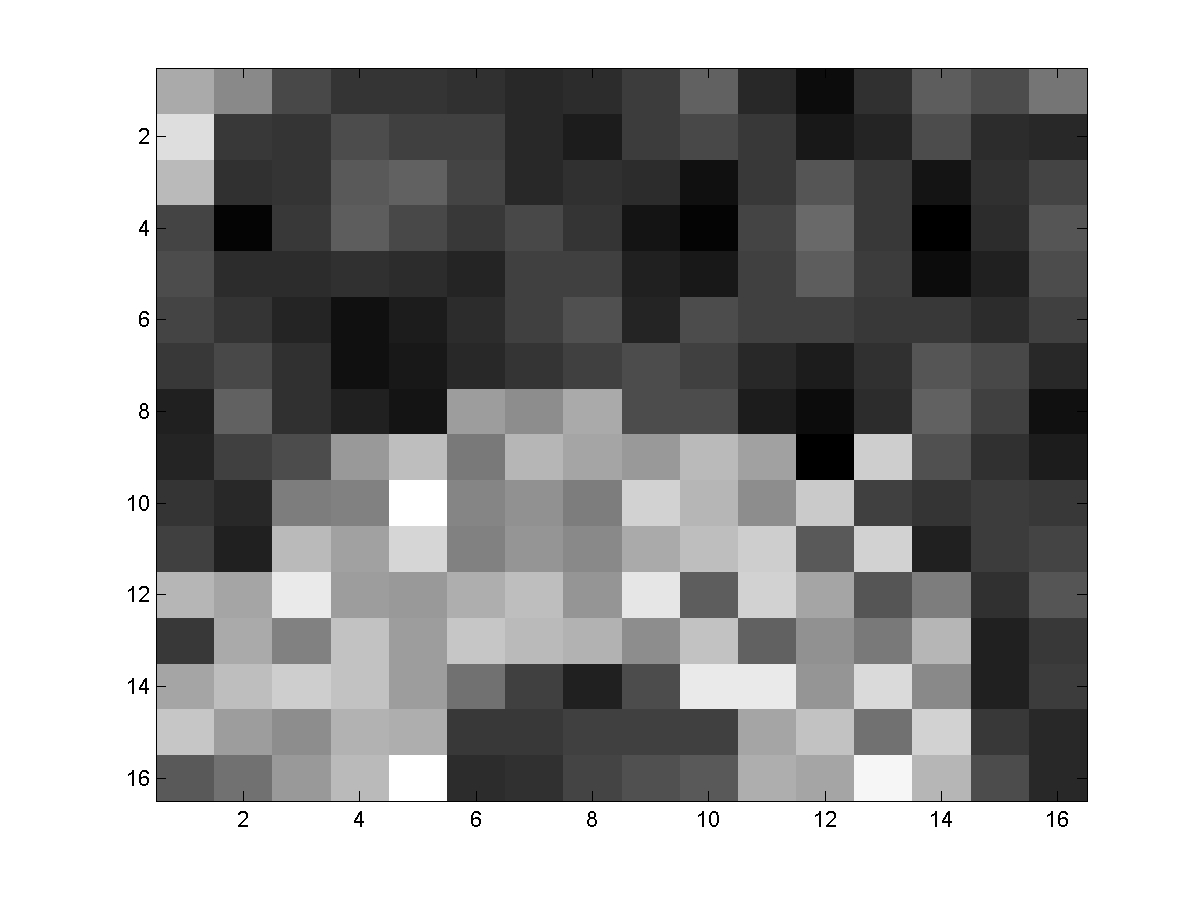}
\includegraphics[height = 40mm, width=40mm]{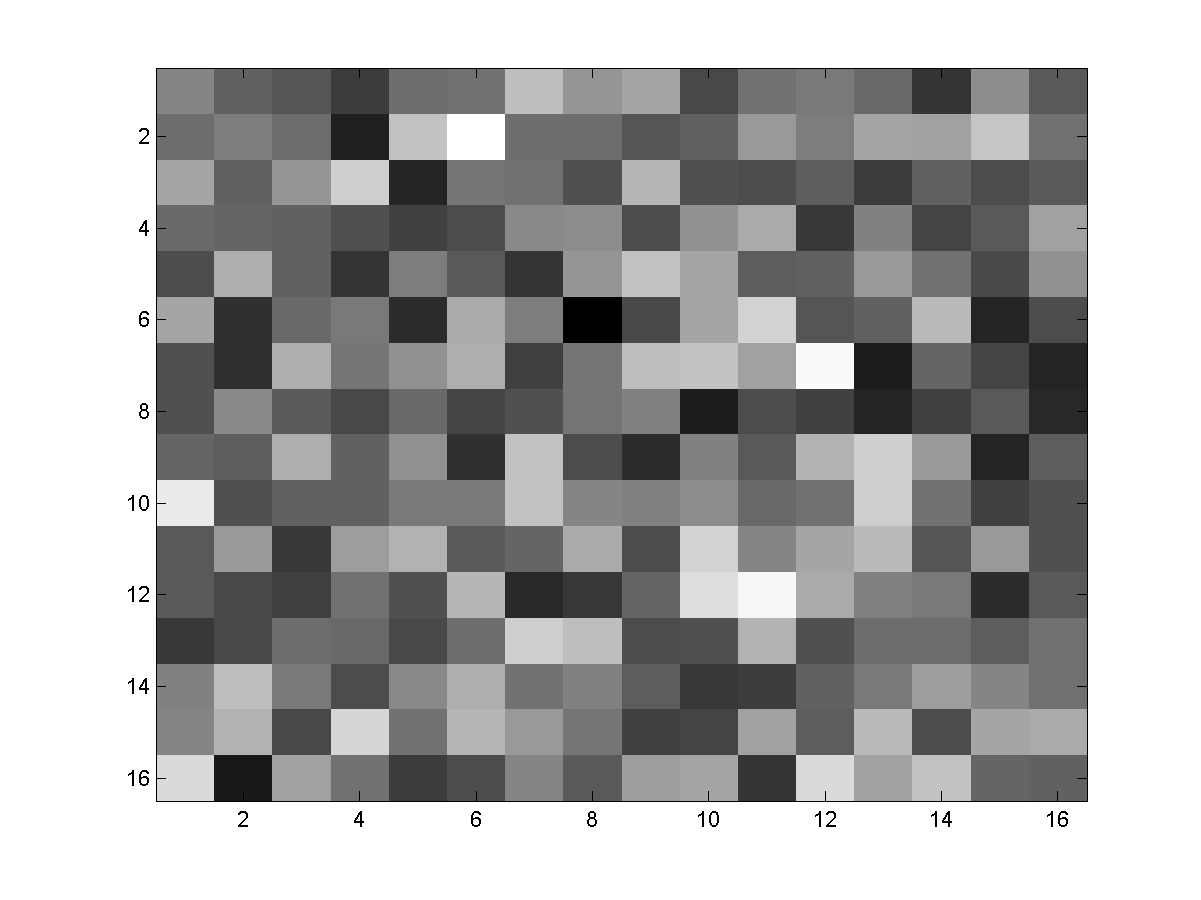}
\caption{\label{fig:sim_sources}Simulated example: the 3 sources.}
\end{figure}

\begin{figure}
\centering
\includegraphics[height=90mm, width=140mm]{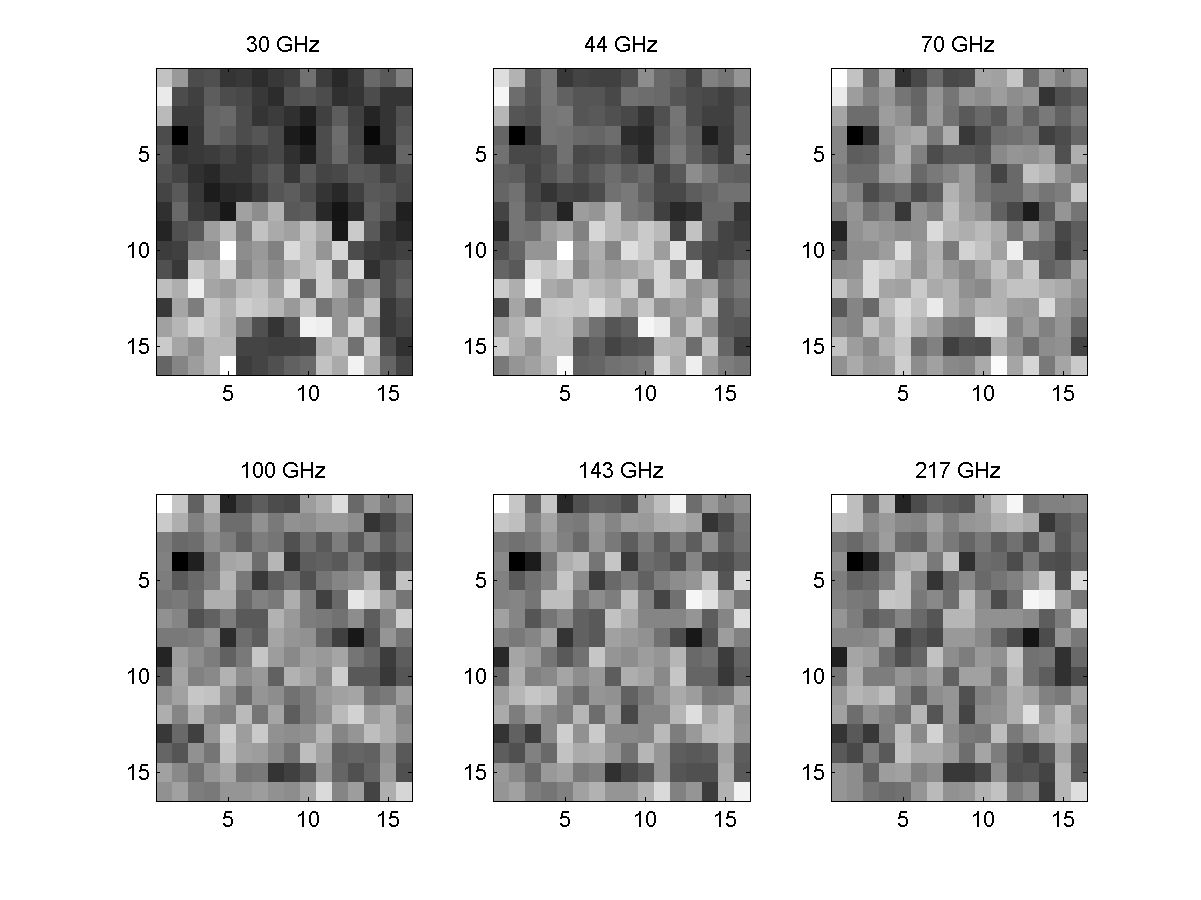}
\caption{\label{fig:sim_data}Simulated example: the 6 observed images.}
\end{figure}

\begin{figure}
\centering
\includegraphics[height = 40mm, width=40mm]{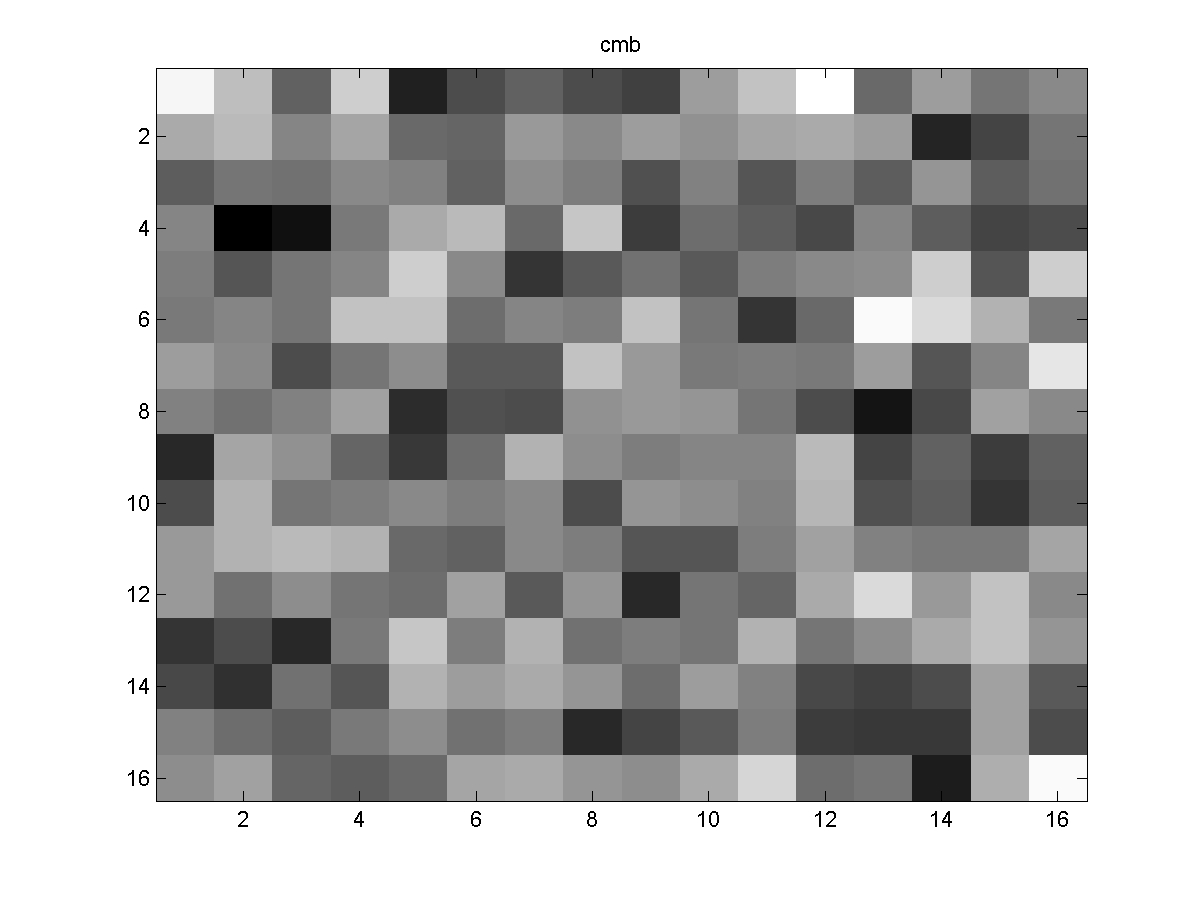}
\includegraphics[height = 40mm, width=40mm]{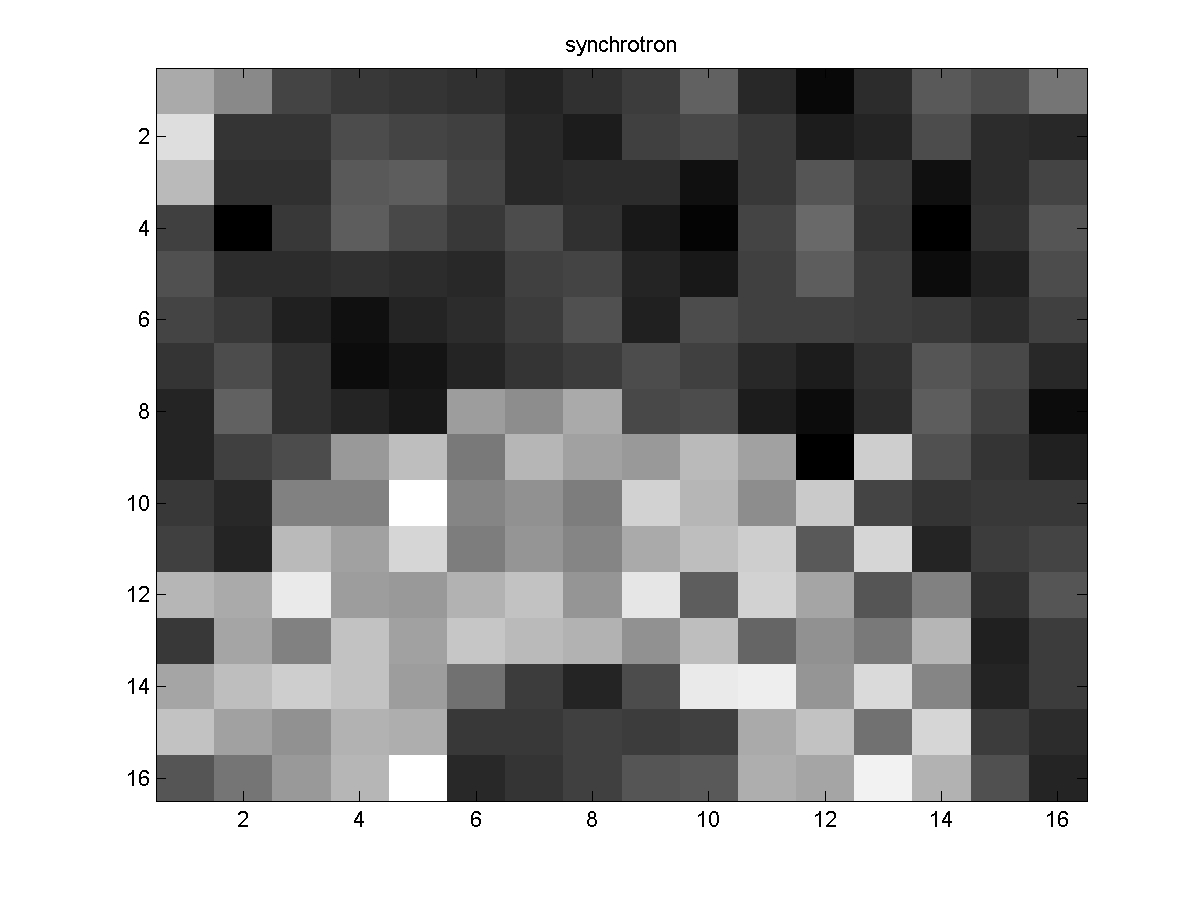}
\includegraphics[height = 40mm, width=40mm]{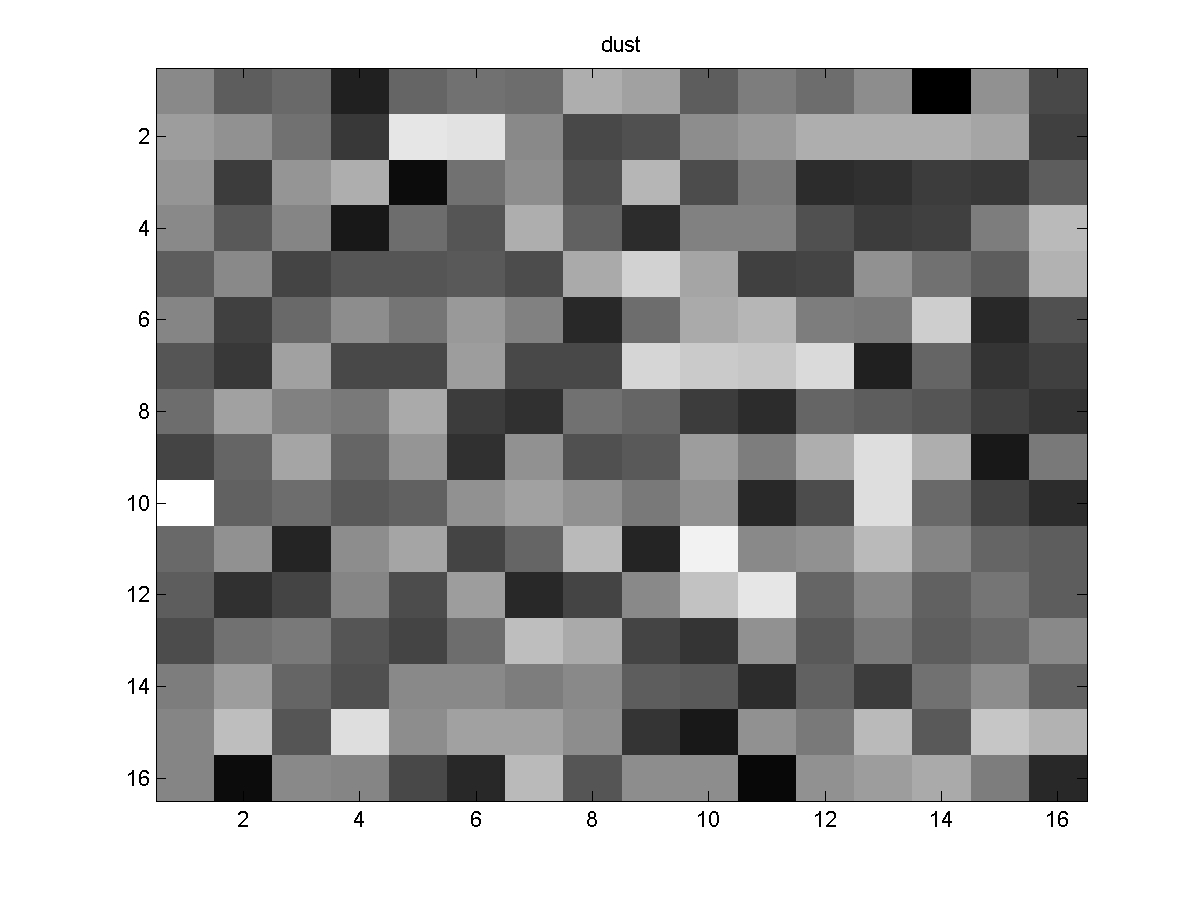}
\caption{\label{fig:sim_results}Simulated example: posterior means of the 3 sources.}
\end{figure}

\begin{figure}
\centering
\begin{tabular}{cccc}
LS          & \includegraphics[scale=0.1]{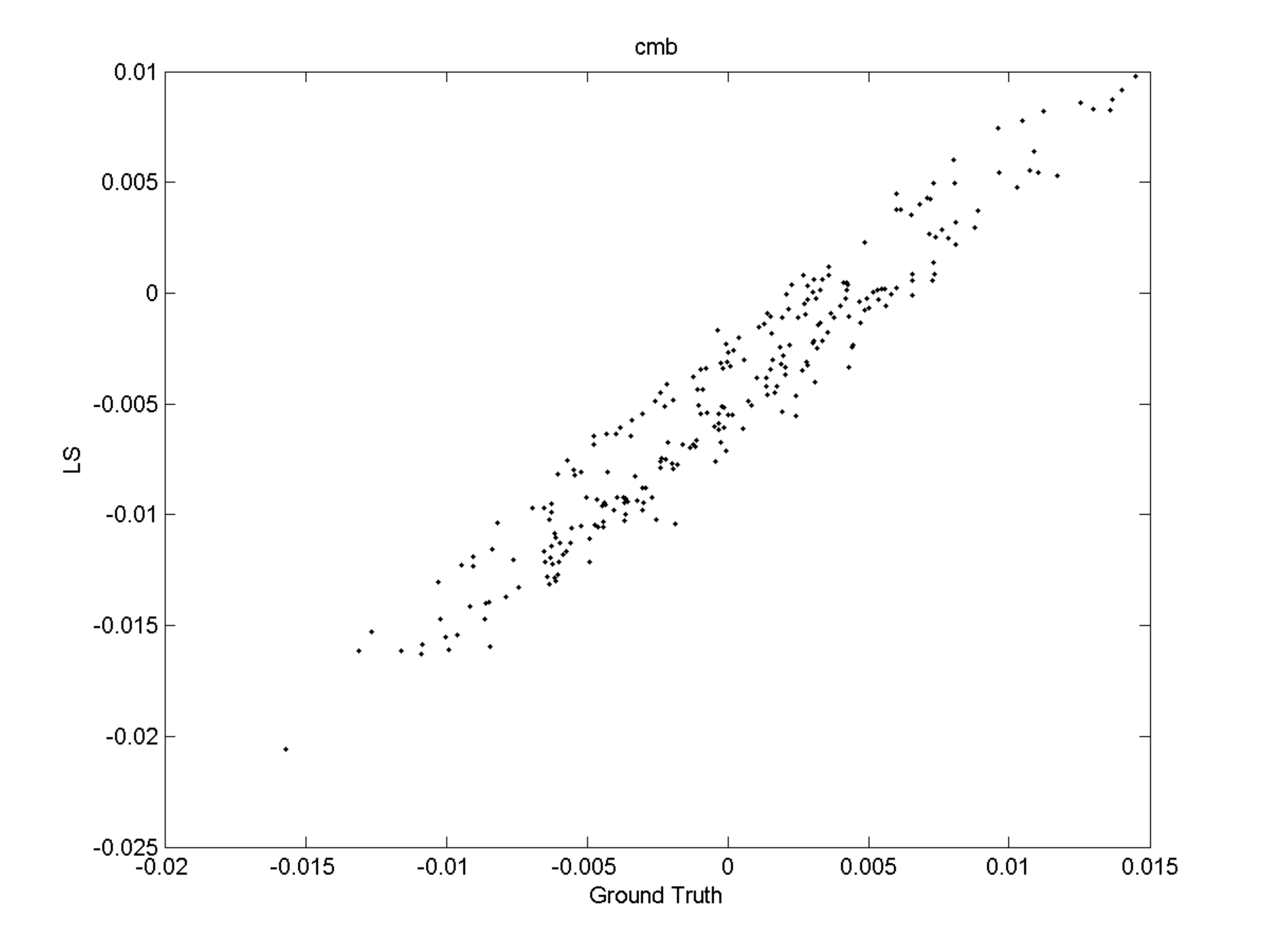} &  \includegraphics[scale=0.1]{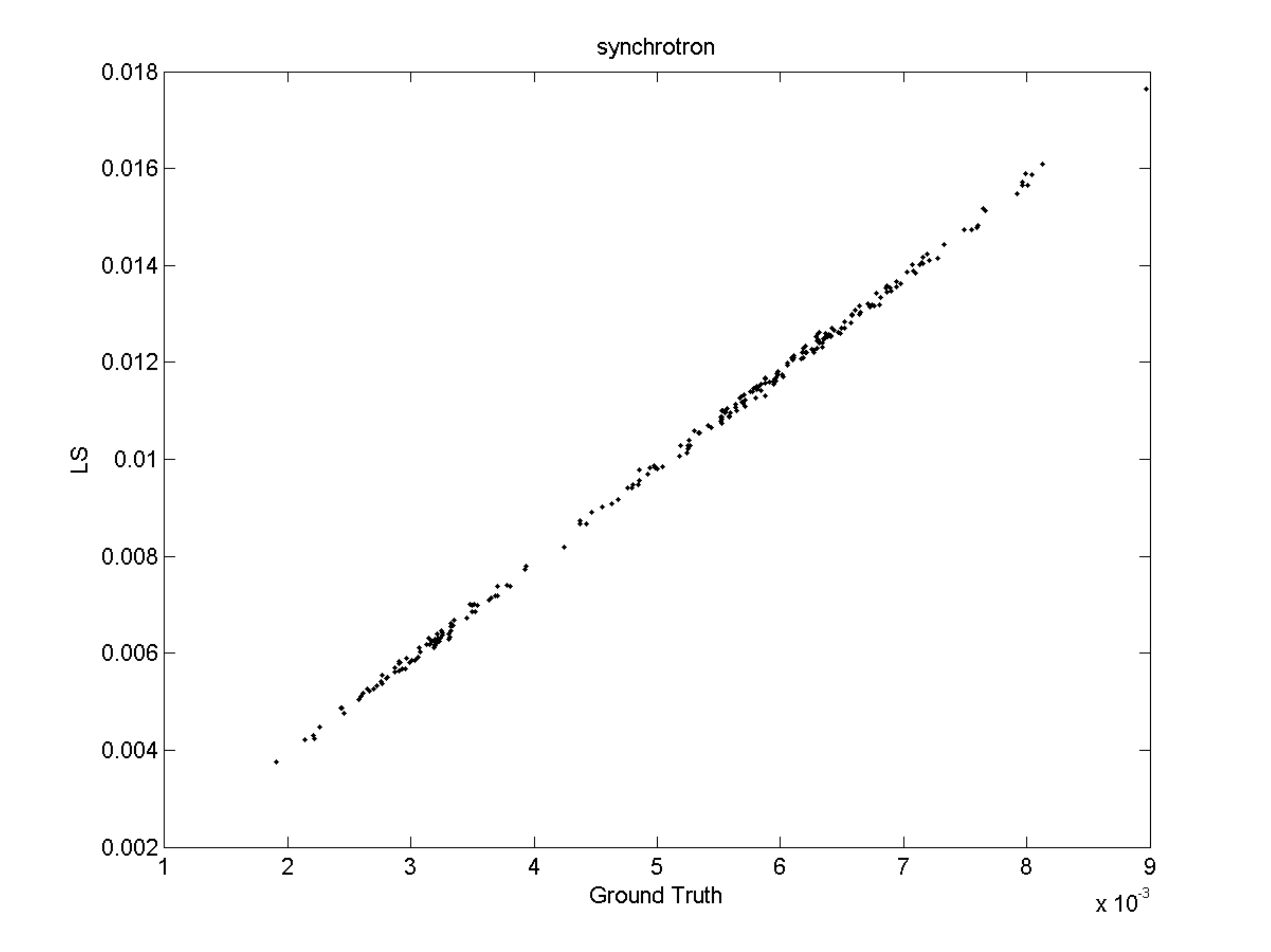} &  \includegraphics[scale=0.1]{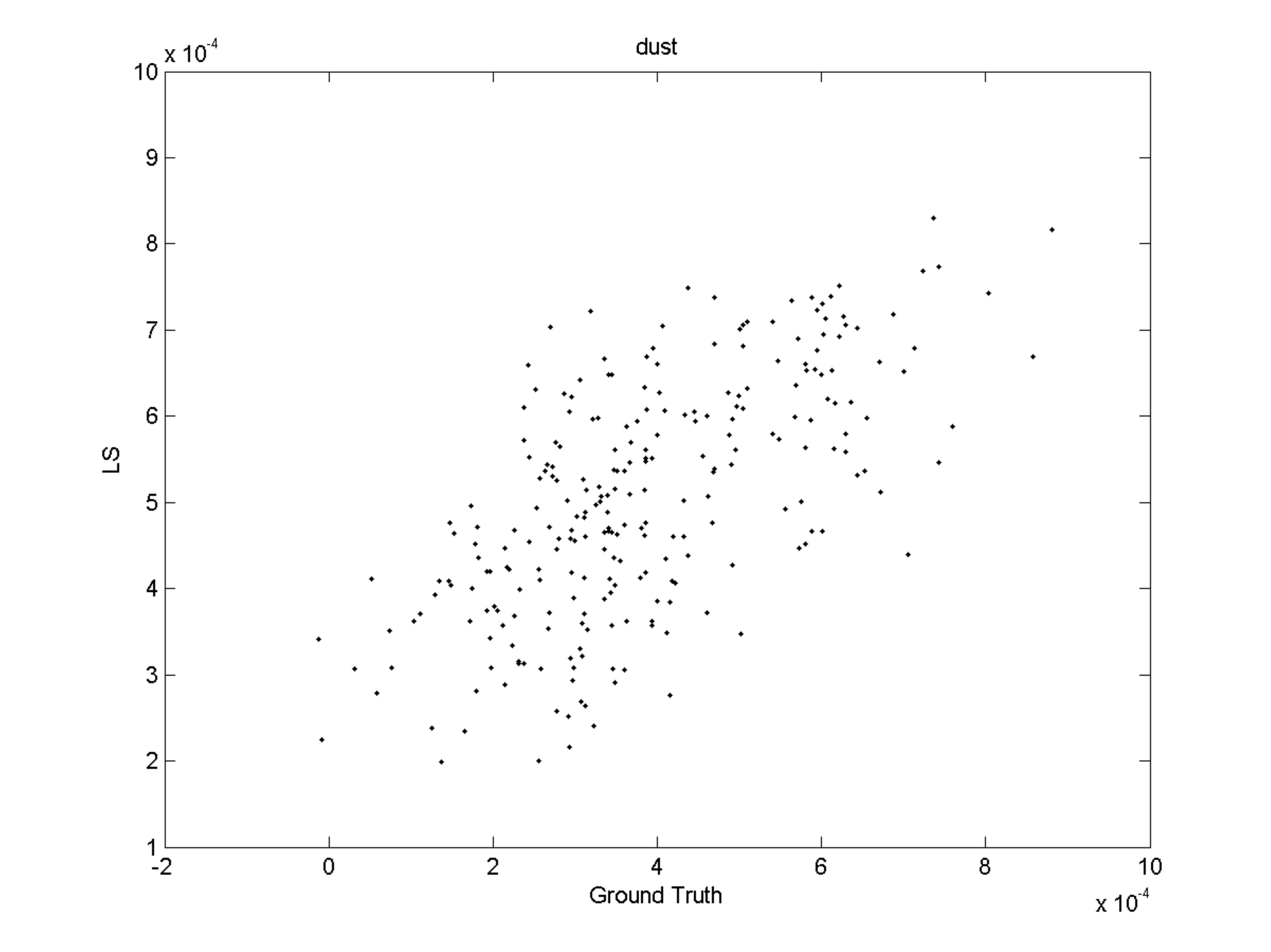} \\
fastICA &  \includegraphics[scale=0.1]{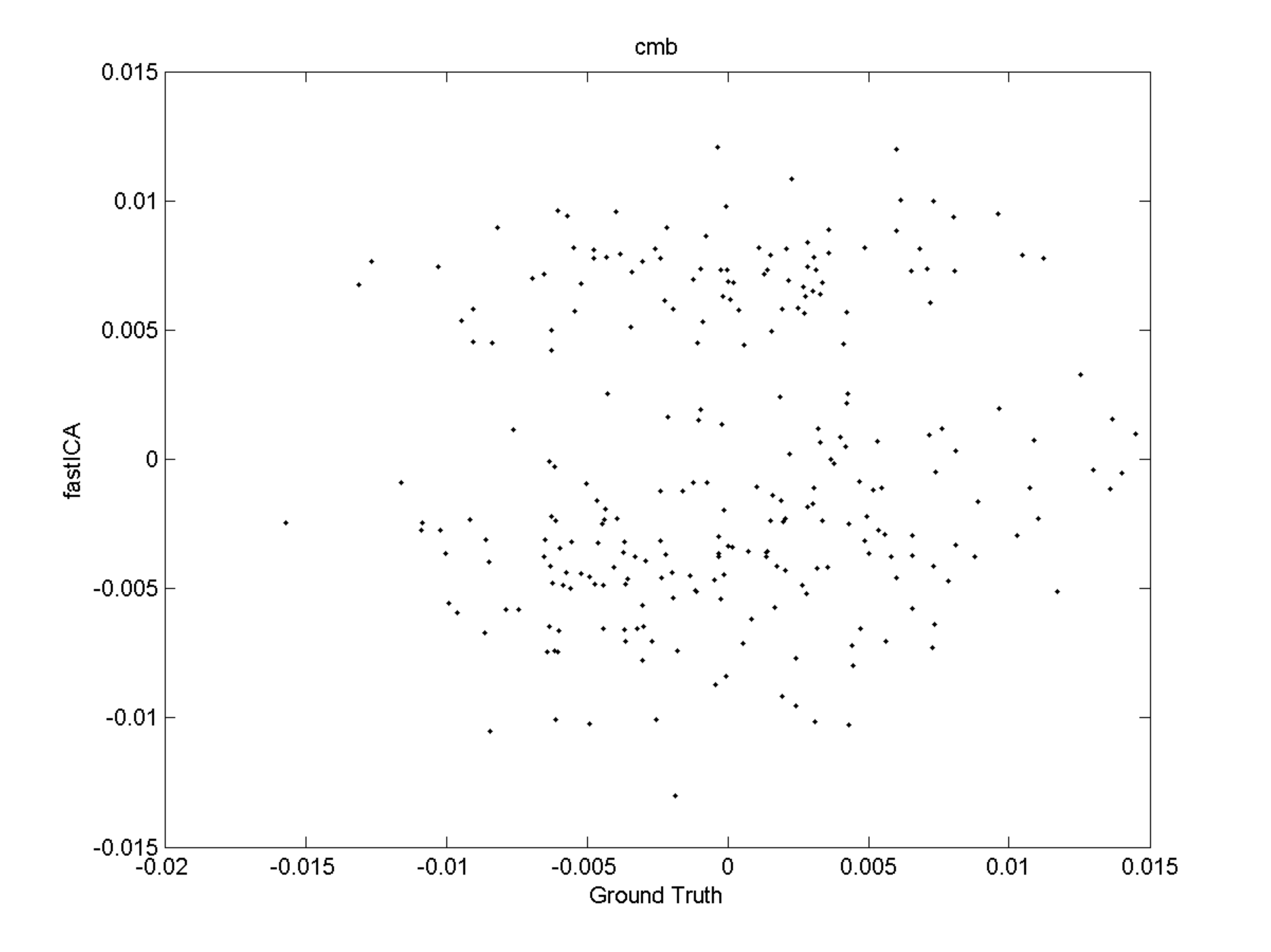} &  \includegraphics[scale=0.1]{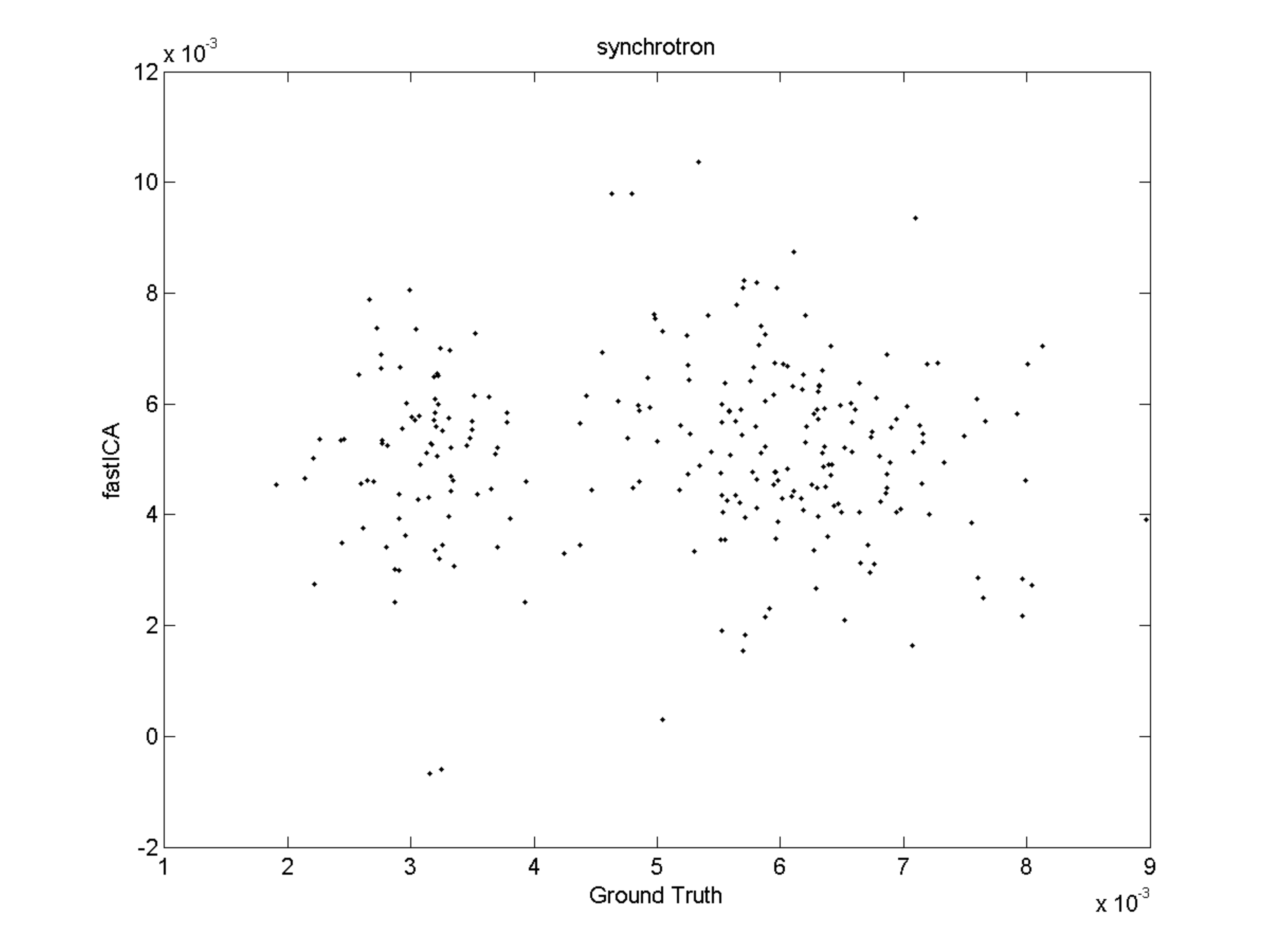} &  \includegraphics[scale=0.1]{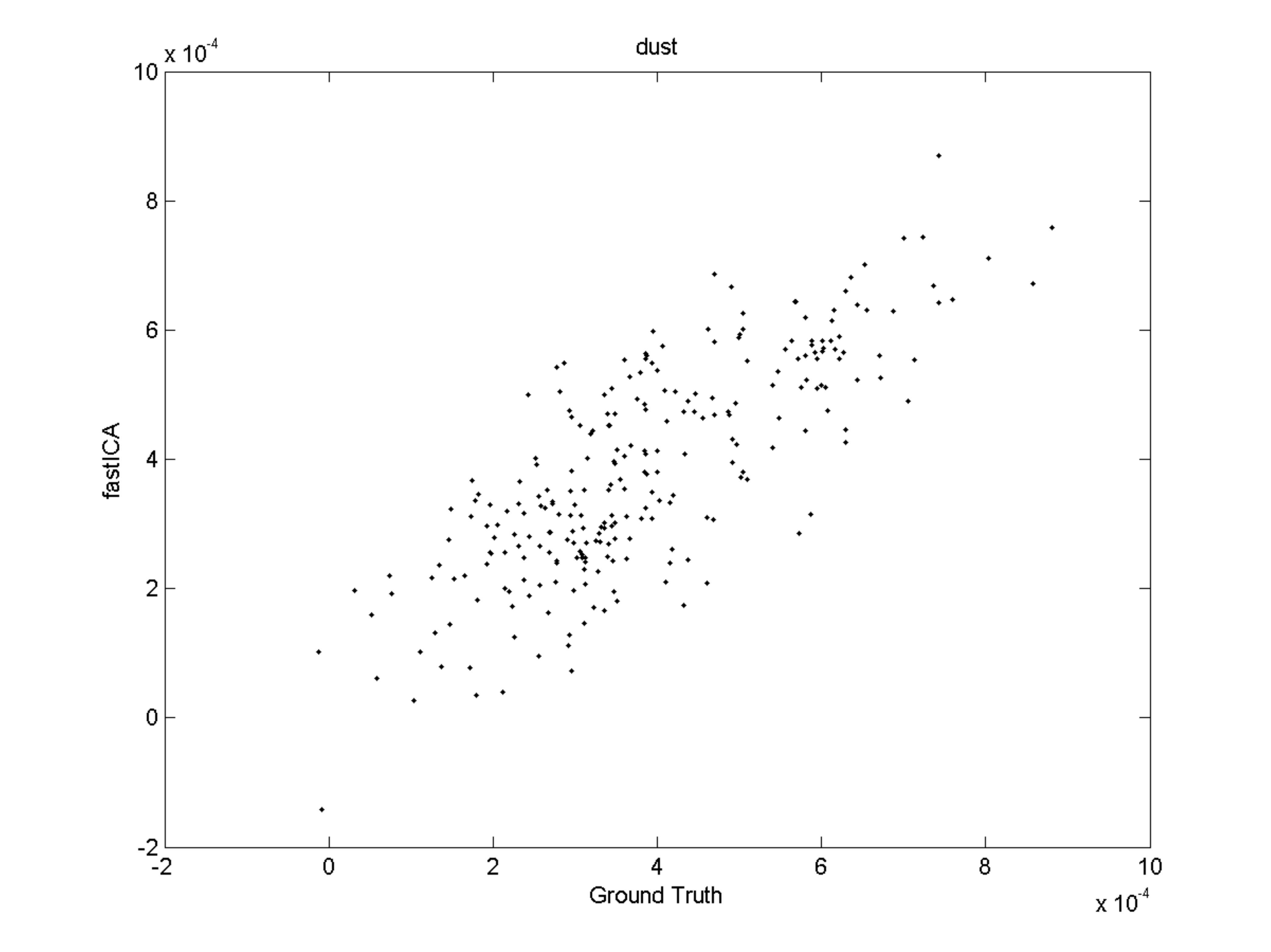} \\
MCMC &  \includegraphics[scale=0.1]{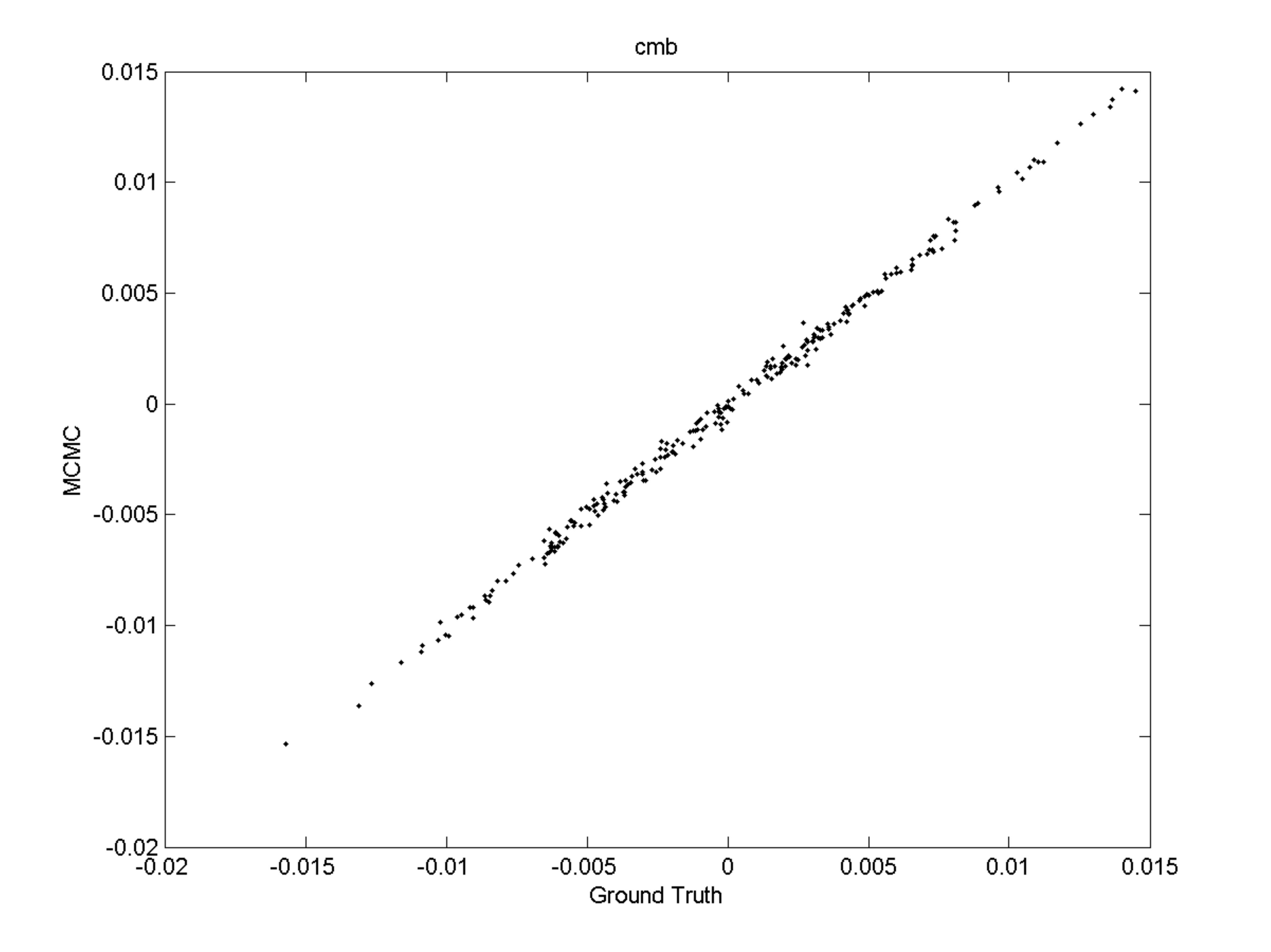} &  \includegraphics[scale=0.1]{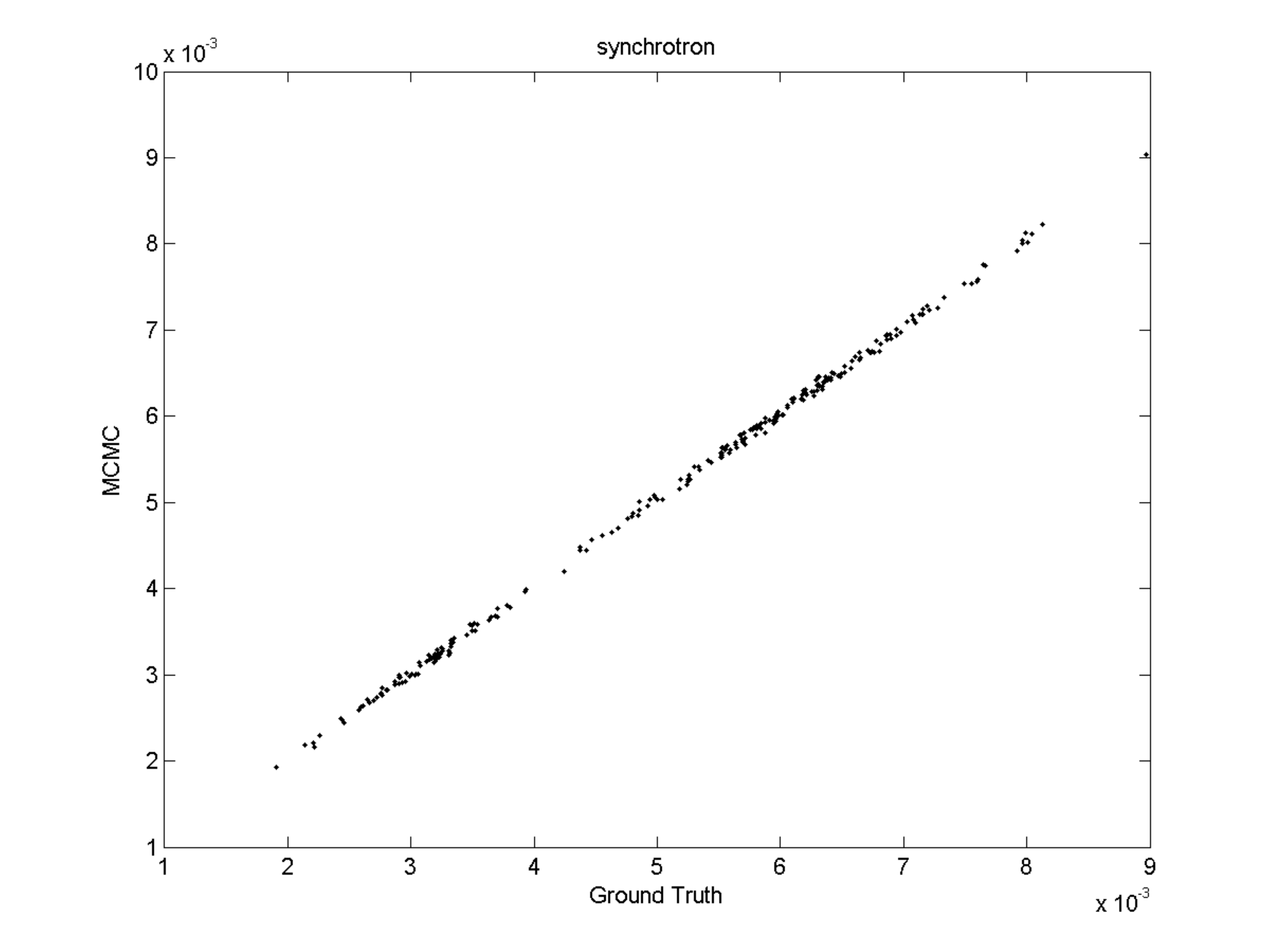} &  \includegraphics[scale=0.1]{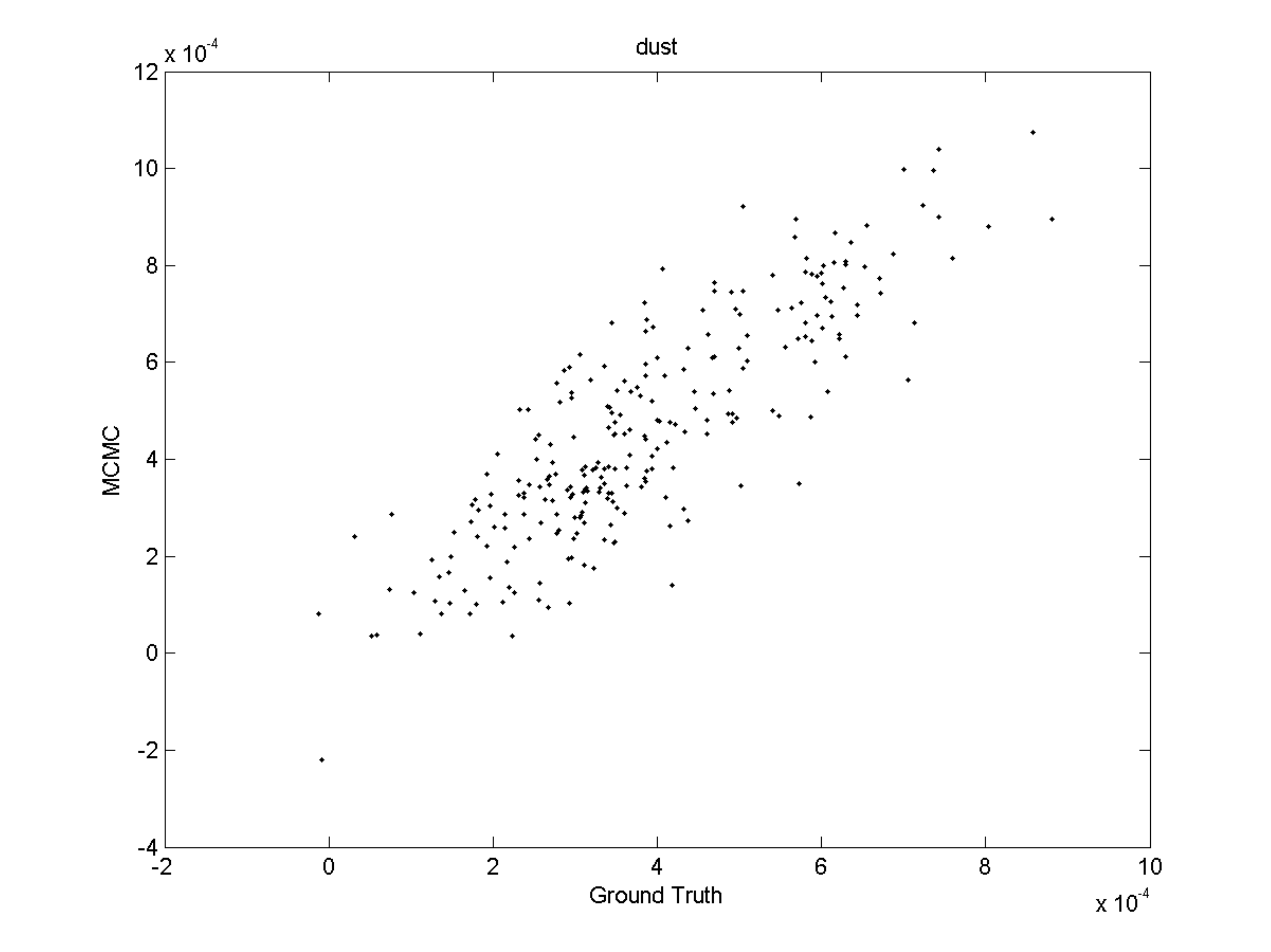} \\
This paper     &  \includegraphics[scale=0.1]{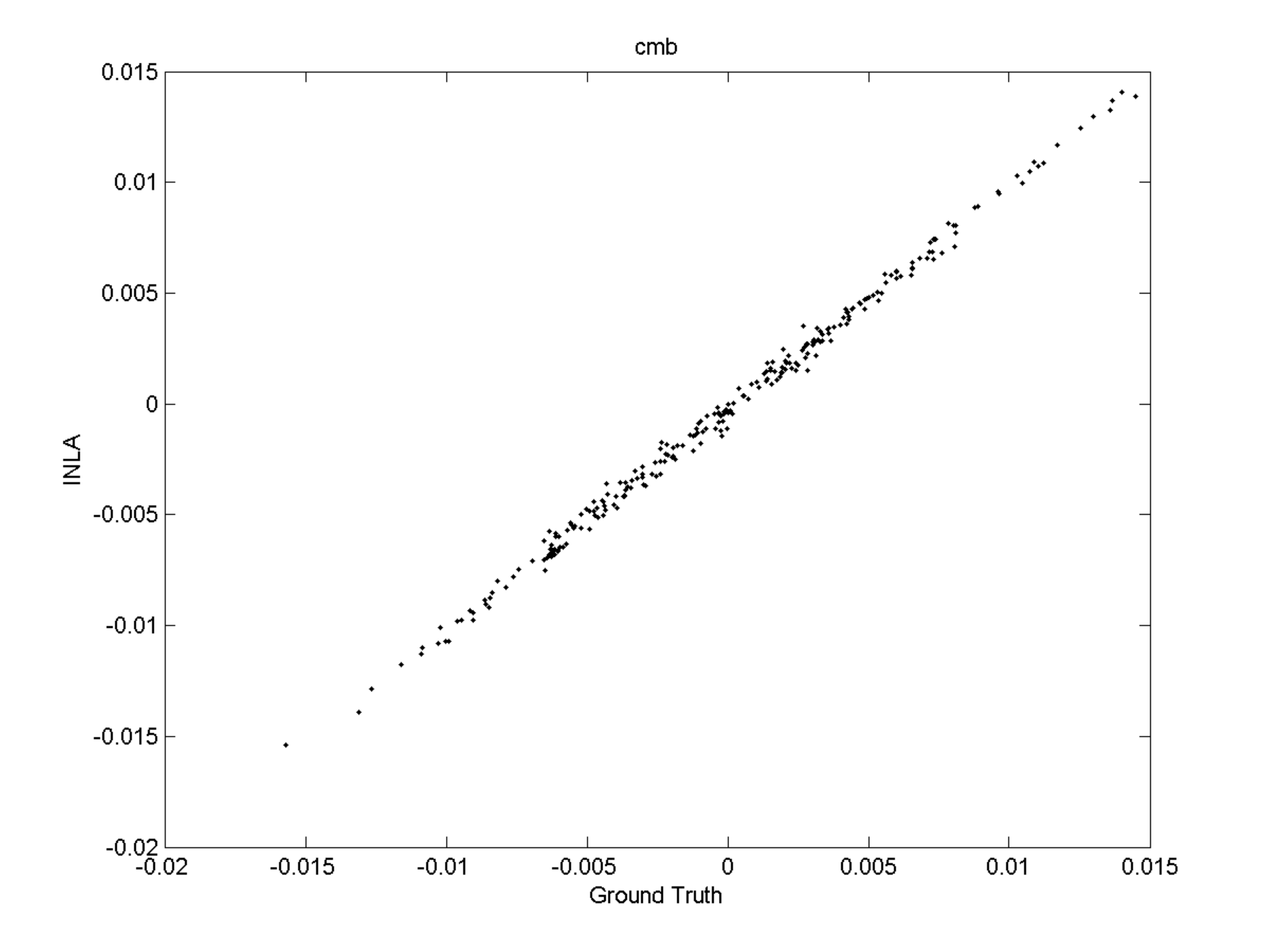} &  \includegraphics[scale=0.1]{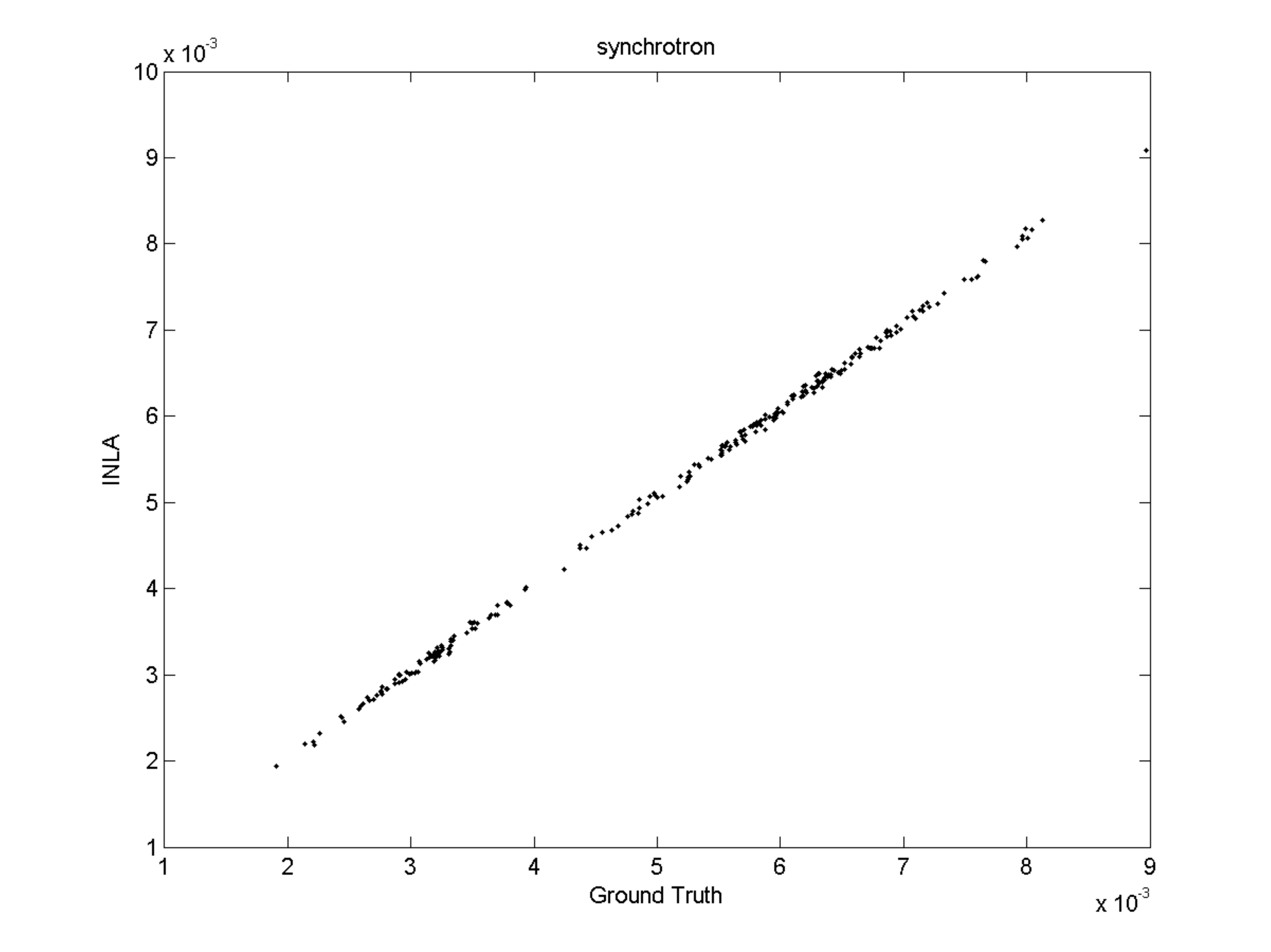} &  \includegraphics[scale=0.1]{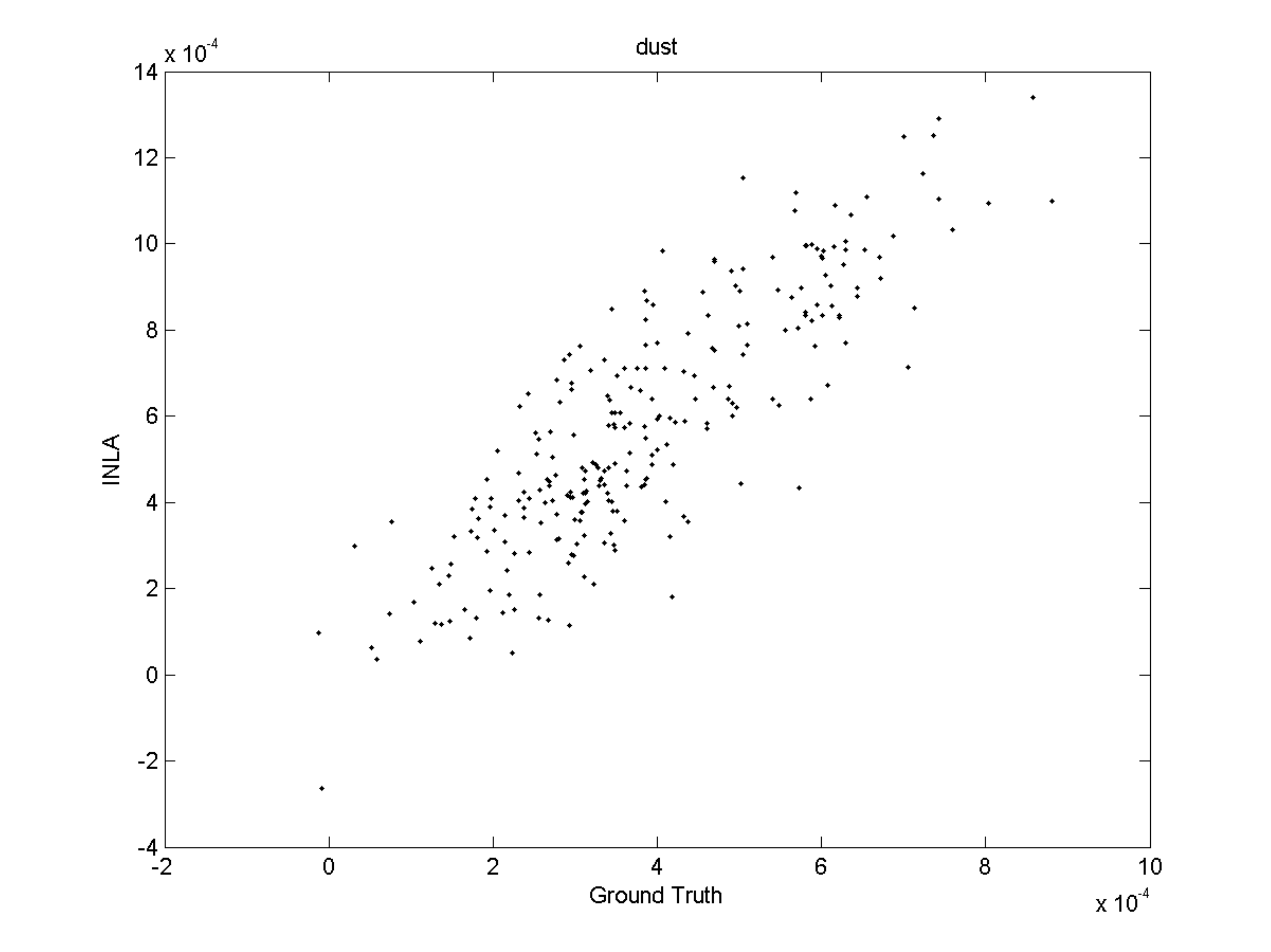} \\
& Source 1 & Source 2 & Source 3 
\end{tabular}
\caption{\label{fig:sim_compare}Scatter plots of observed versus posterior means for the 3 sources with 4 algorithms: least squares, fast ICA, Bayesian implemented by MCMC and Bayesian implemented by the approach of this paper.}
\end{figure}

As an illustration, 3 sources were mixed into 6 components according to the matrix
\[ \bm{A} = \left( \begin{array}{ccc} 
1.26 &  29.11 &   0.20 \\
   1.22 &   9.96 &   0.34 \\
   1.14 &   2.71 &   0.63 \\
   1.00 &   1.00 &   1.00 \\
   0.78 &   0.37 &   1.55 \\
   0.43 &   0.11 &   2.51
    \end{array} \right), \]
which corresponds to the mixing components of CMB, synchrotron and galactic dust respectively, as described in Section \ref{sec:model}, at 30, 44, 70, 100, 143 and 217  GHz, with $\theta_{s}=-2.8$ and $\theta_{d}=1.4$, normalised so that the 4th row is made of ones.   Figure \ref{fig:sim_sources} and \ref{fig:sim_data} shows the ground truth and data respectively.  The prior distributions  for $\theta_s$, $\theta_d$ and $\phi_i$ follow those described in \ref{sec:prior}.  This problem is sufficiently small ($n_s = 3$, $J = 256$) to be solved without blocking.  The parameter vector $\bm{\Phi}$ is of dimension 6 with the grid ${\cal Q}$ composed of about 50,000 points.  Figure \ref{fig:sim_results} shows the posterior means of the 3 sources calculated via Eq.\ \ref{eq:inla_postmean}.
As a comparison with other common methods of source separation, in Figure \ref{fig:sim_compare} are scatter plots of true versus estimated source pixel values via standard least squares and fast ICA, as well as Bayesian inference implemented by MCMC and the approach of this paper. We see that the Bayesian method produces the most accurate result for either implementation, but it is noted that the approach of this paper is substantially faster than MCMC.

\section{Analysis of 7 year WMAP data}
\label{sec:wmap}
\begin{figure}
\centering
\includegraphics[scale=0.22]{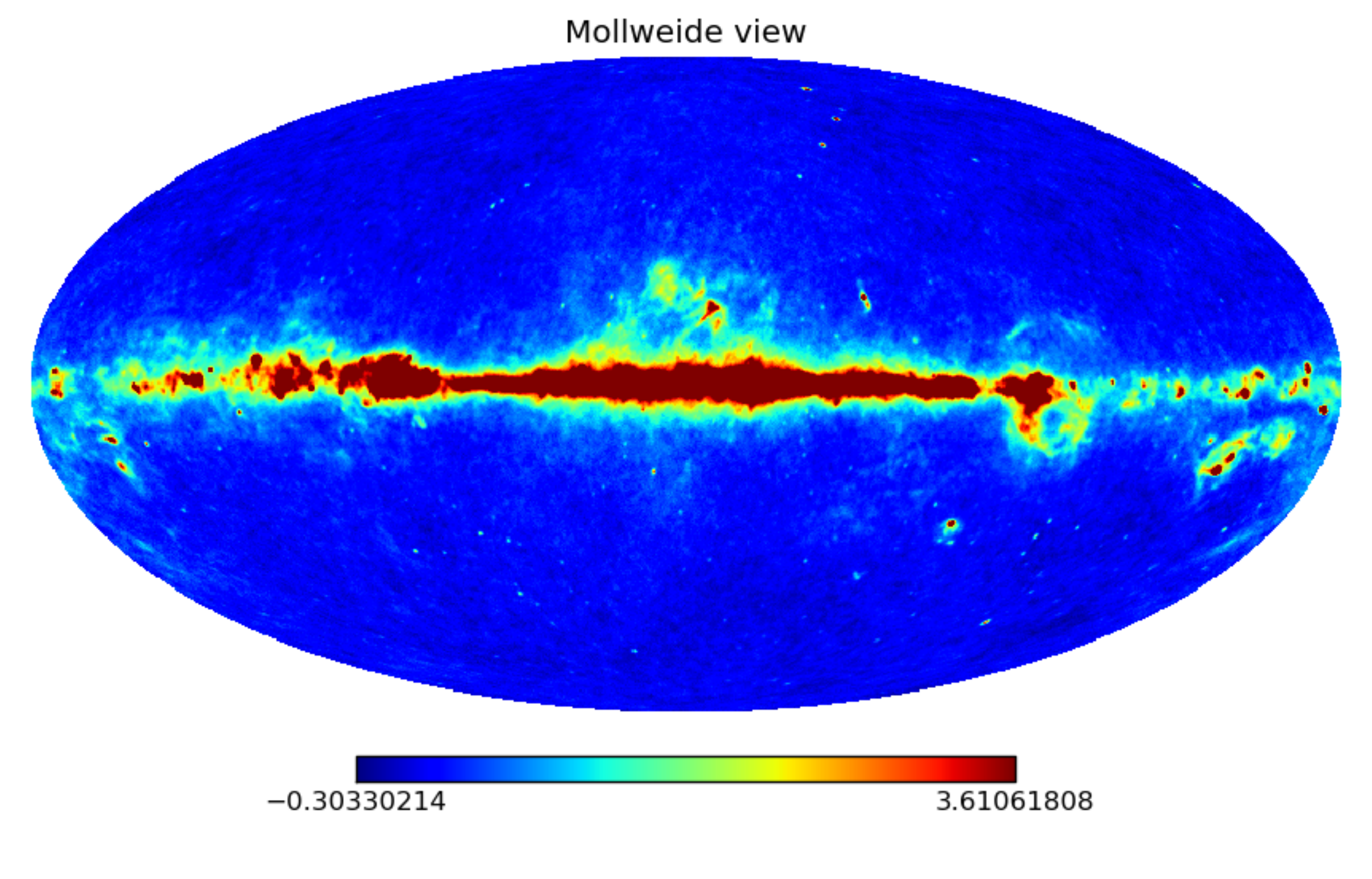}
\includegraphics[scale=0.22]{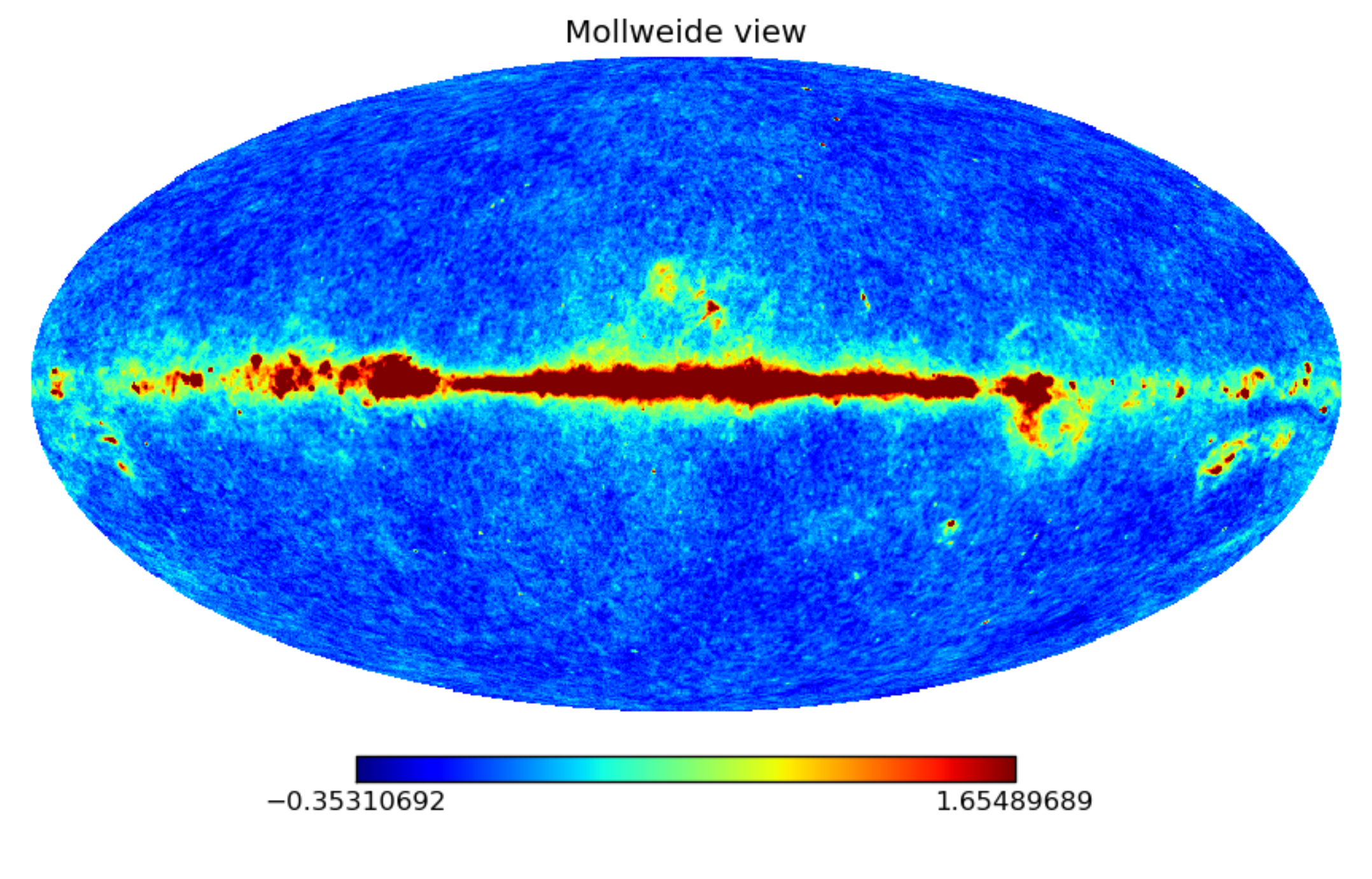}
\includegraphics[scale=0.22]{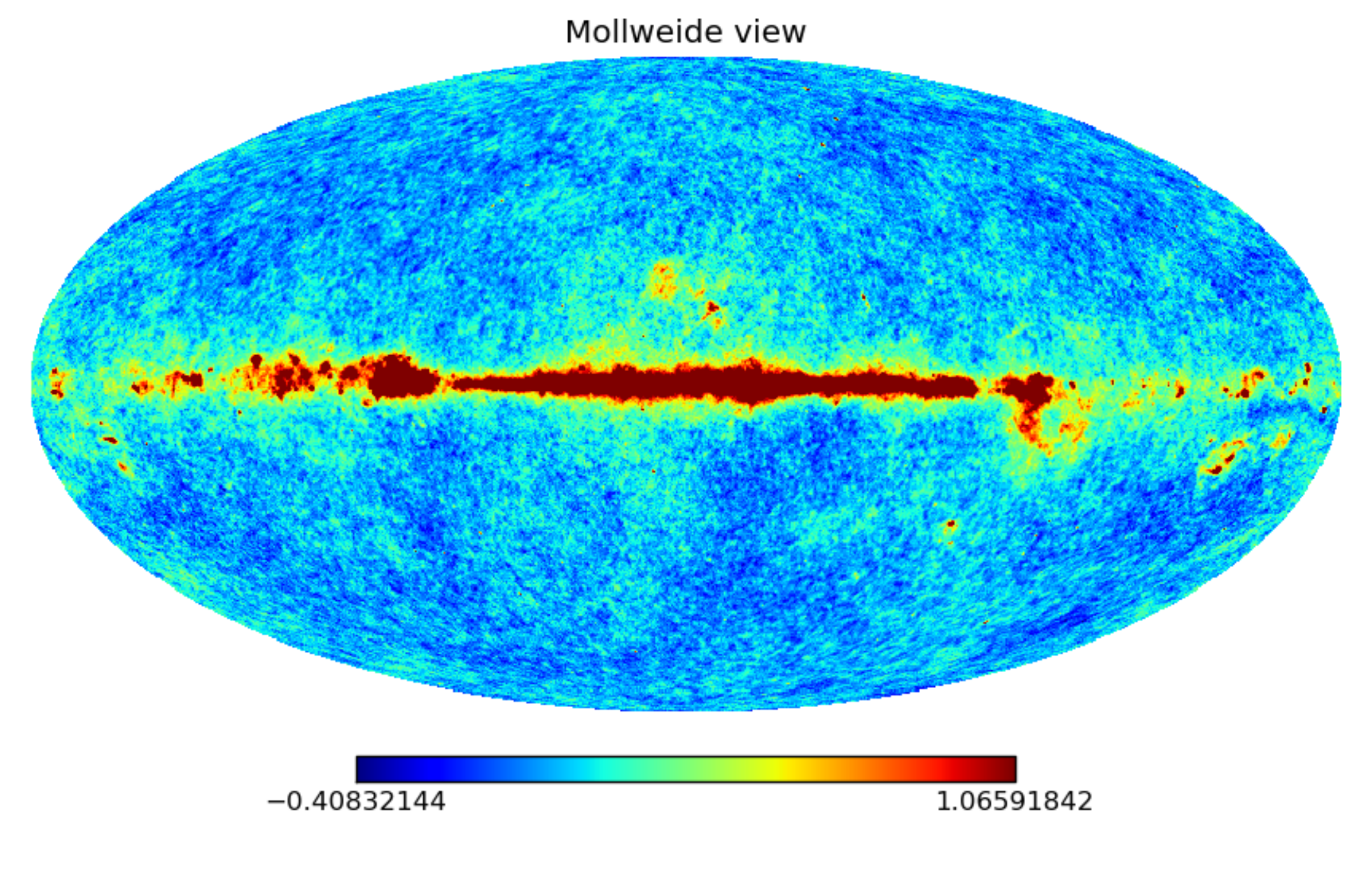} \\
\includegraphics[scale=0.22]{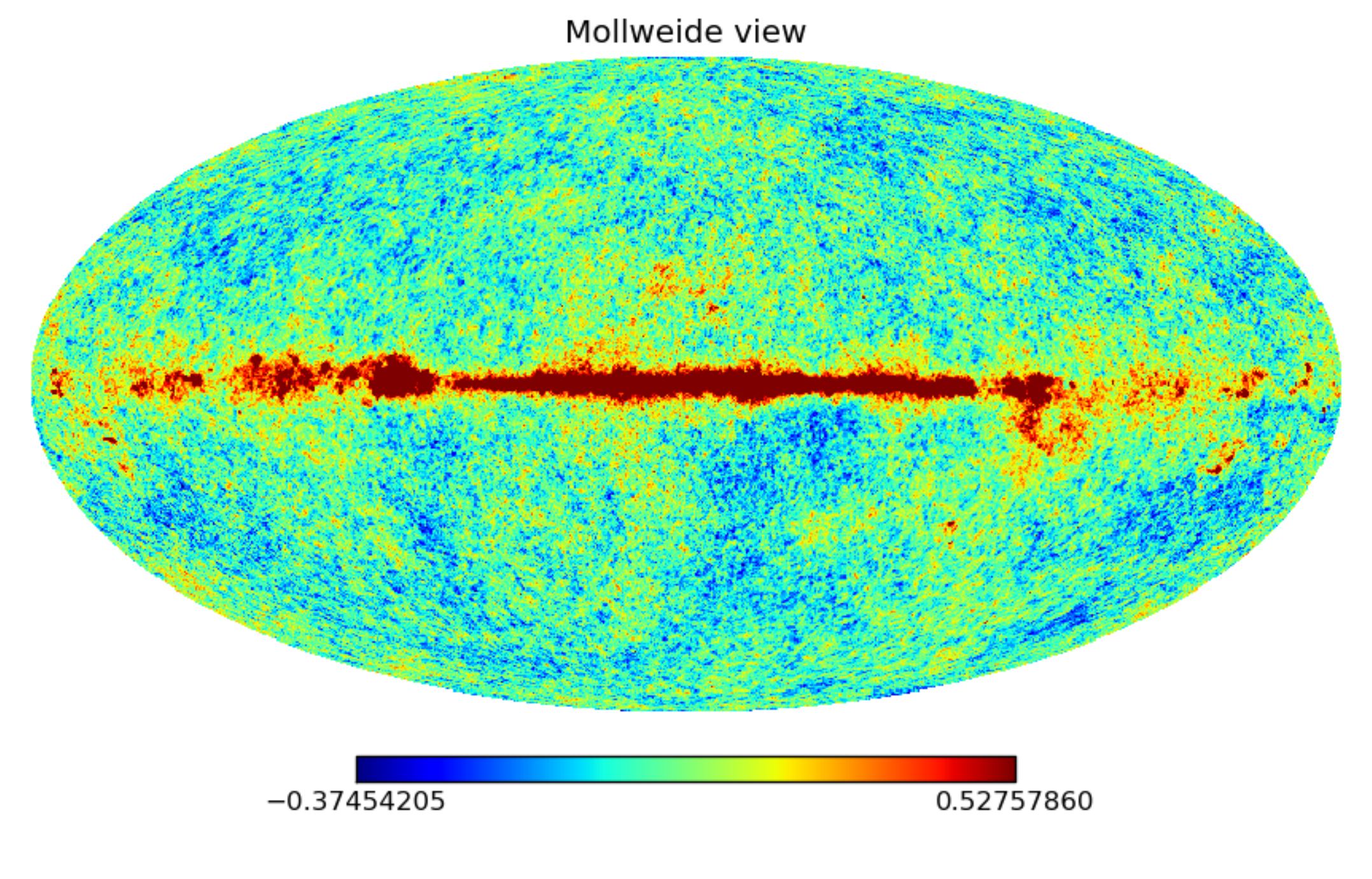}
\includegraphics[scale=0.22]{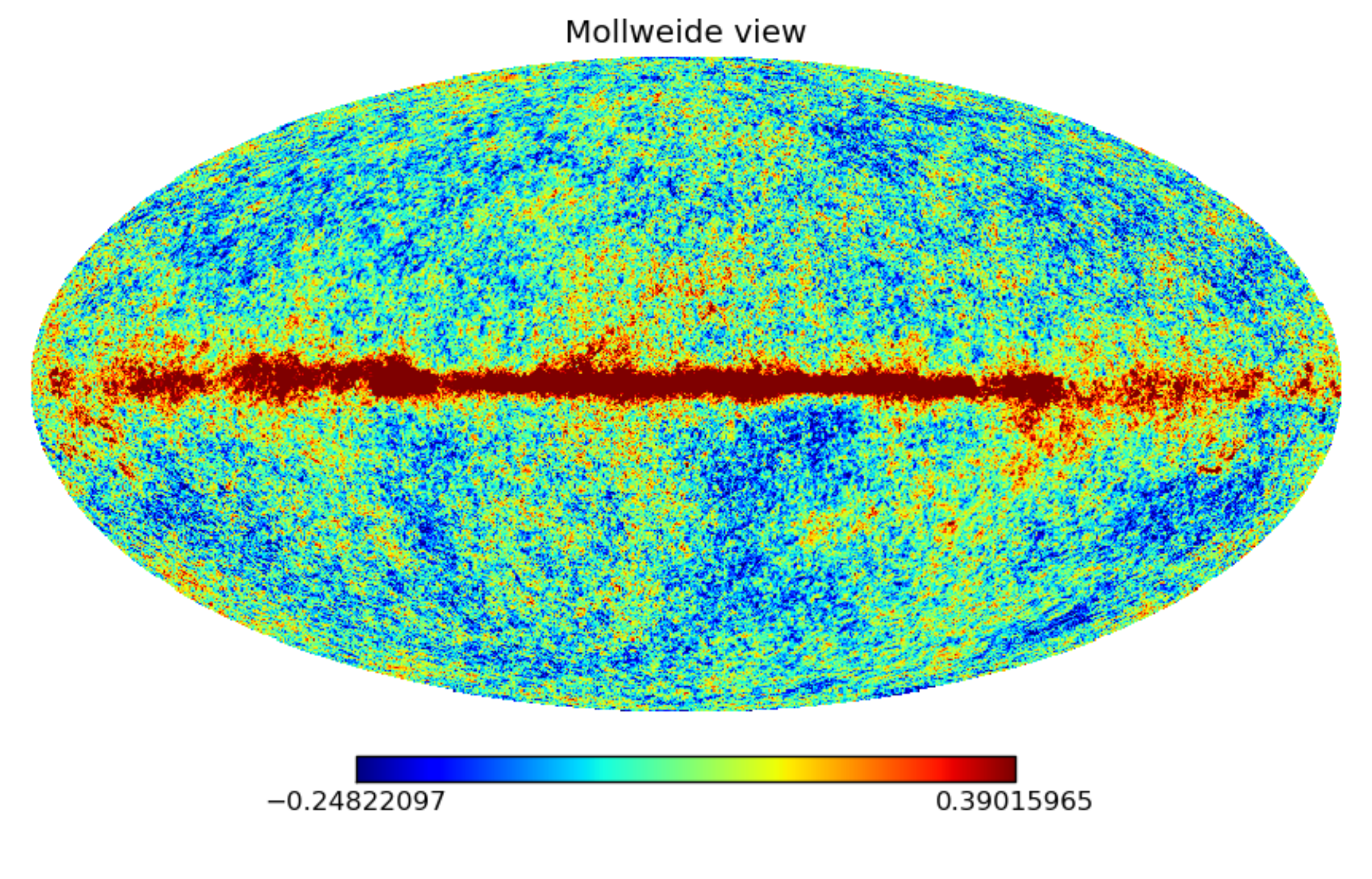}
\caption{\label{fig:wmap}The 7 year WMAP data.}
\end{figure}

The seven year WMAP data was analysed using the procedure of Section \ref{sec:postcalc}.  WMAP data consist of 5 images of $J = 3 \times 2^{20} = 3,145,728$ pixels (see Figure \ref{fig:wmap}) which were divided into 6144 blocks of $512$ pixels for the analysis.  

Separation into the 4 sources described in Section \ref{sec:model}, following the method described in Sections \ref{sec:prior} and \ref{sec:postcalc}, was implemented.  The parameter vector $\bm{\Psi} = (\theta_d,\theta_s,\phi_1,\ldots,\phi_4)$ has dimension 6 and, for the computation of $q_W(\bm{\Psi} \, | \, \bm{Y}_W)$, the grid ${\cal Q}$ was computed as described in Section \ref{sec:postcalc} which led to a grid size of at most 50,000 points and sometimes much smaller. On the same PC as was used for Section \ref{sec:simdata}, the computation of each $p_W(\bm{\Psi} \, | \, \bm{Y}_W)$ and $\mathbb{E}(\bm{S}_W \, | \, \bm{Y}_W)$ through Eqs.\ \ref{eq:inla_p_constant} and \ref{eq:inla_postmean} took about 40 seconds with MATLAB code.  On a single processor, this equates to about 72 hours to complete the full map. The most time-consuming operation was the Cholesky decomposition used to compute $\bm{Q}_W^*(\bm{\Psi})^{-1}$. It is noted that processing of different blocks can be done in parallel, so there is great potential to reduce the total computation time if more processors are available.

\begin{figure}
\centering
\includegraphics[scale=0.22]{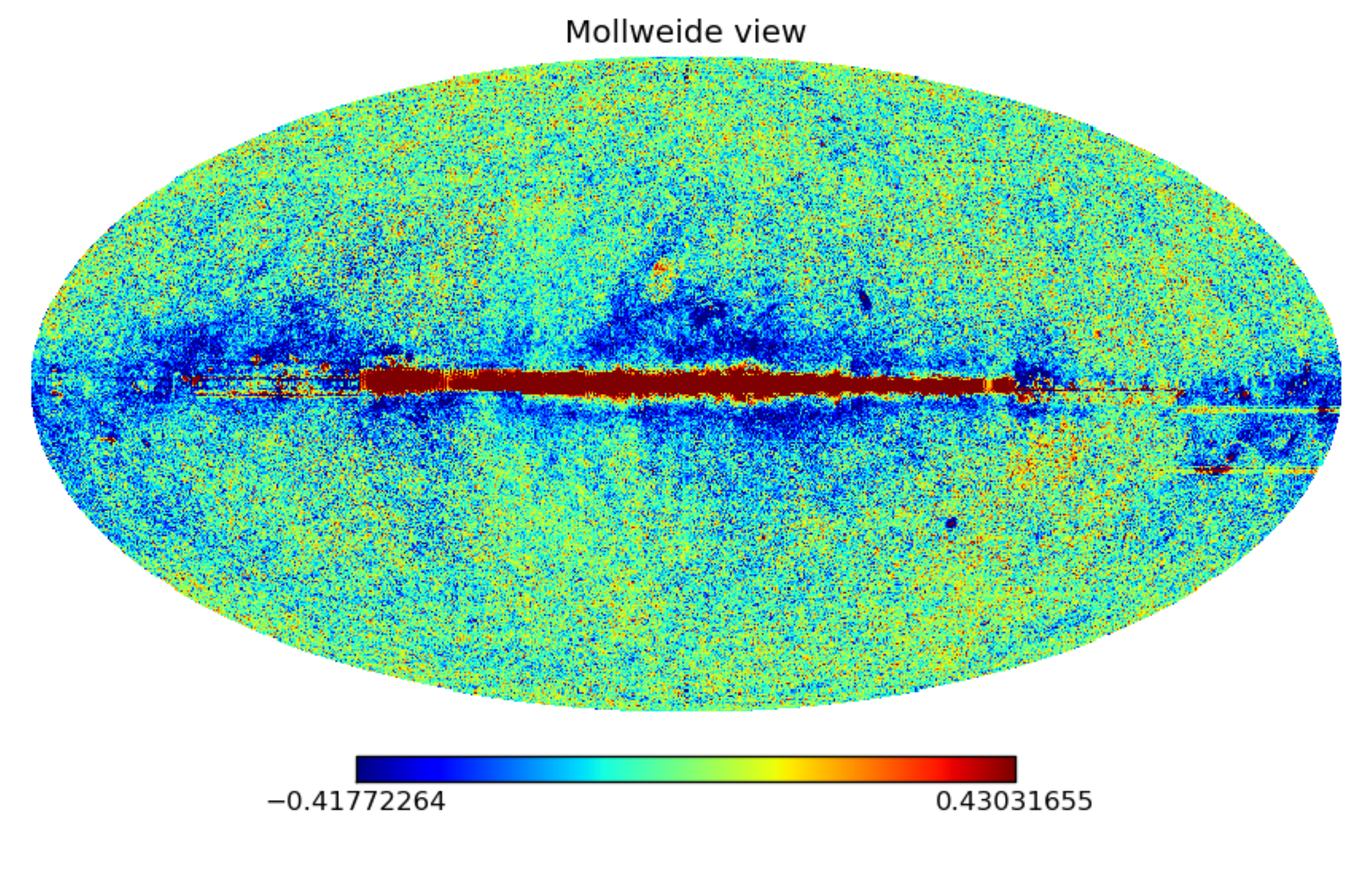}
\includegraphics[scale=0.22]{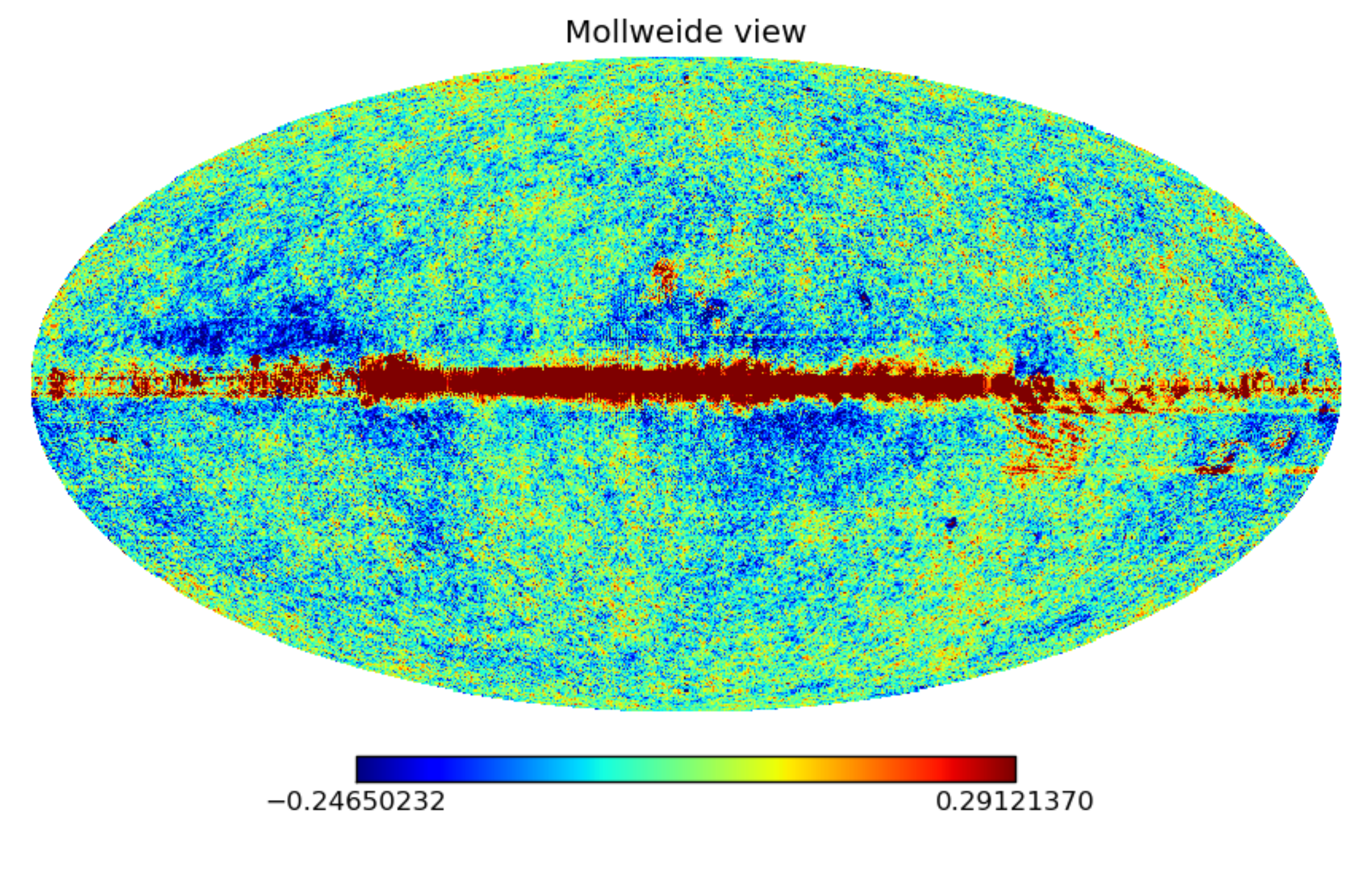}
\includegraphics[scale=0.22]{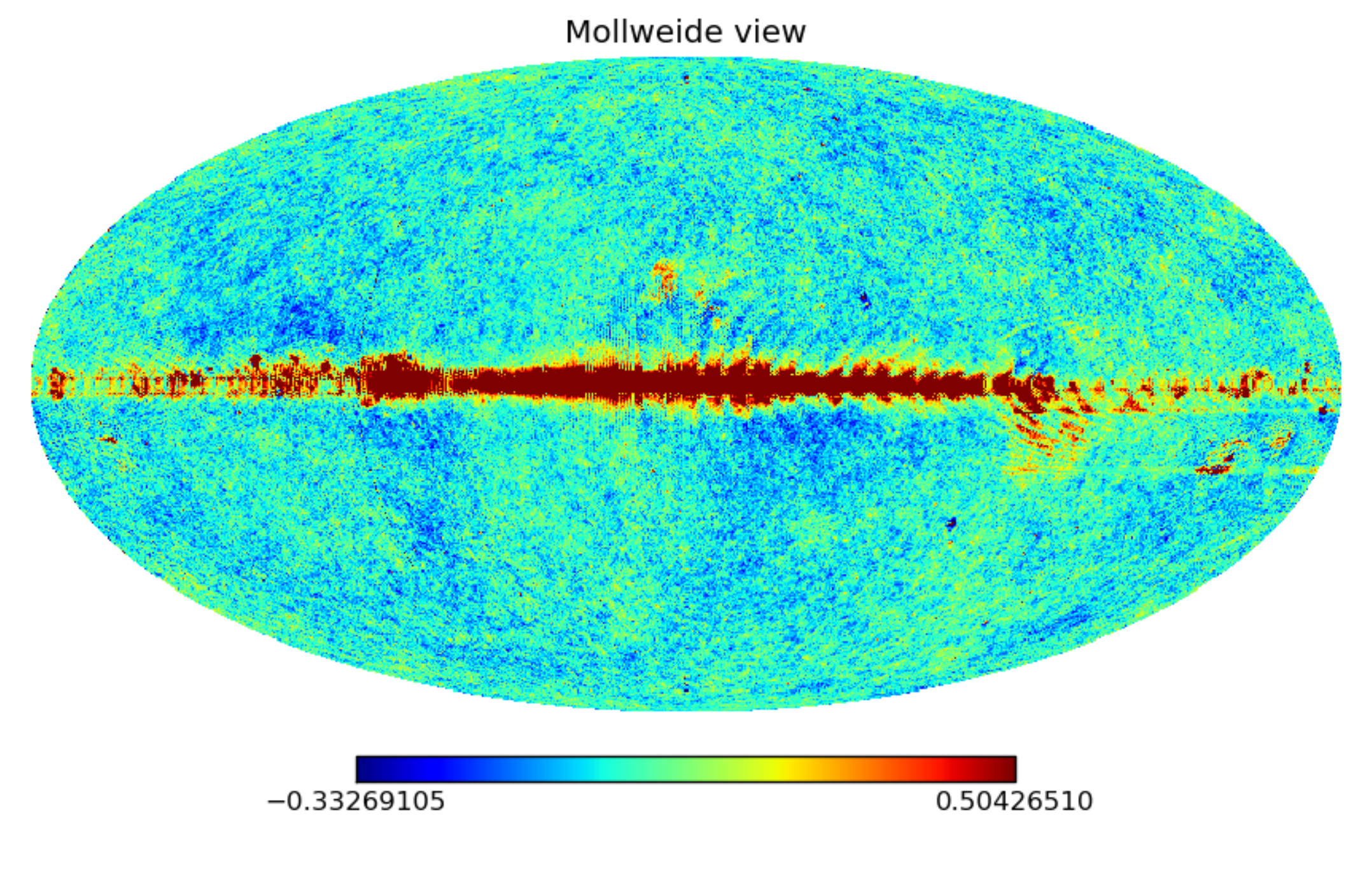} 
\caption{\label{fig:cmb}Posterior means of CMB for WMAP data from Eq.\ \ref{eq:inla_postmean} with (from left to right) a prior mean for the spatial smoothness parameters $\phi_i$ of 1, 5 and 10.}
\end{figure}

\begin{figure}
\centering
\includegraphics[scale=0.22]{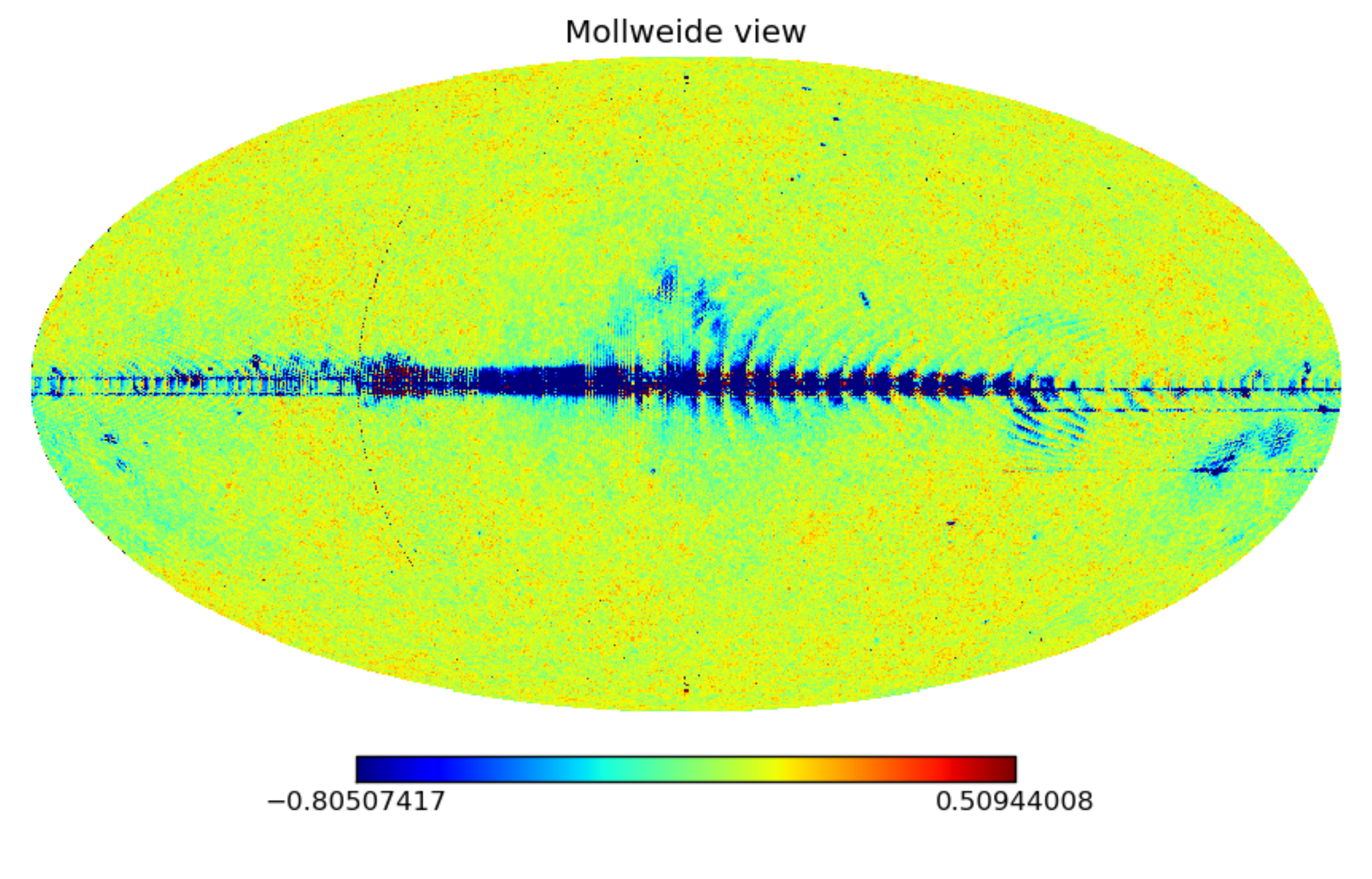}
\includegraphics[scale=0.22]{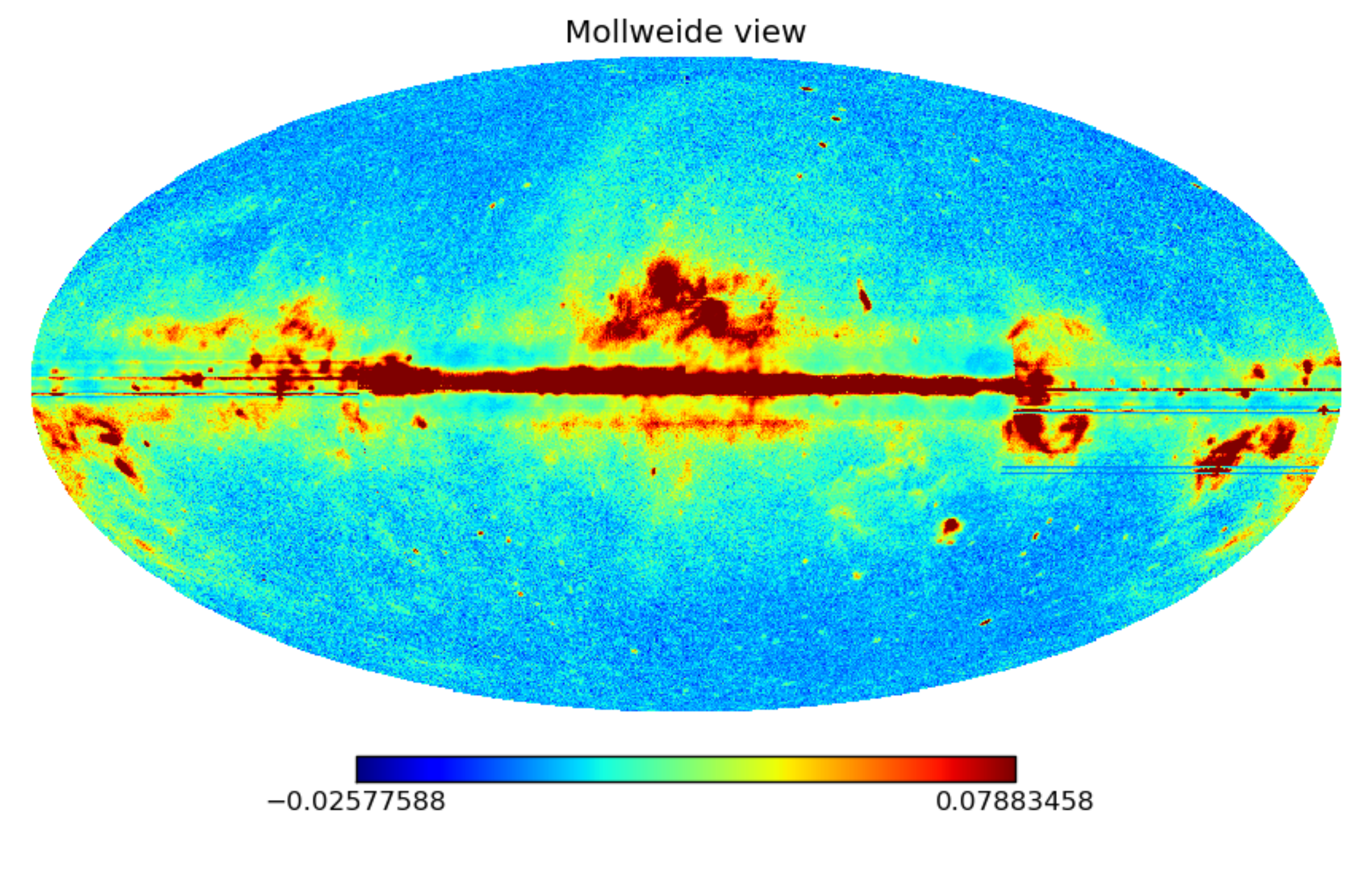}
\includegraphics[scale=0.22]{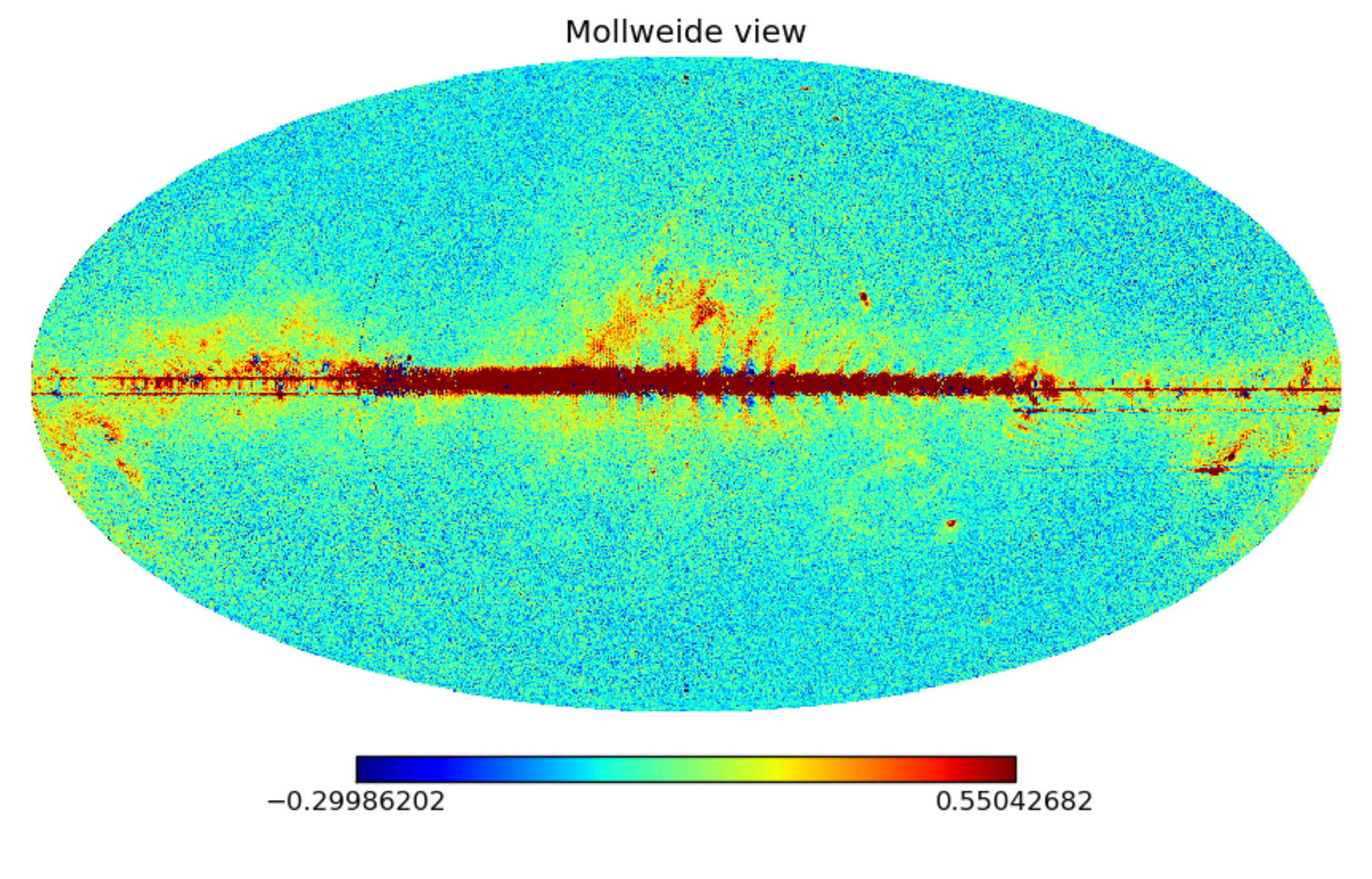} 
\caption{\label{fig:other}Posterior means of (from left to right) synchrotron, galactic dust and free-free emission for WMAP data from Eq.\ \ref{eq:inla_postmean} with  a prior mean for the spatial smoothness parameters $\phi_i$ of 10.}
\end{figure}

\begin{figure}
\centering
\includegraphics[scale=0.6]{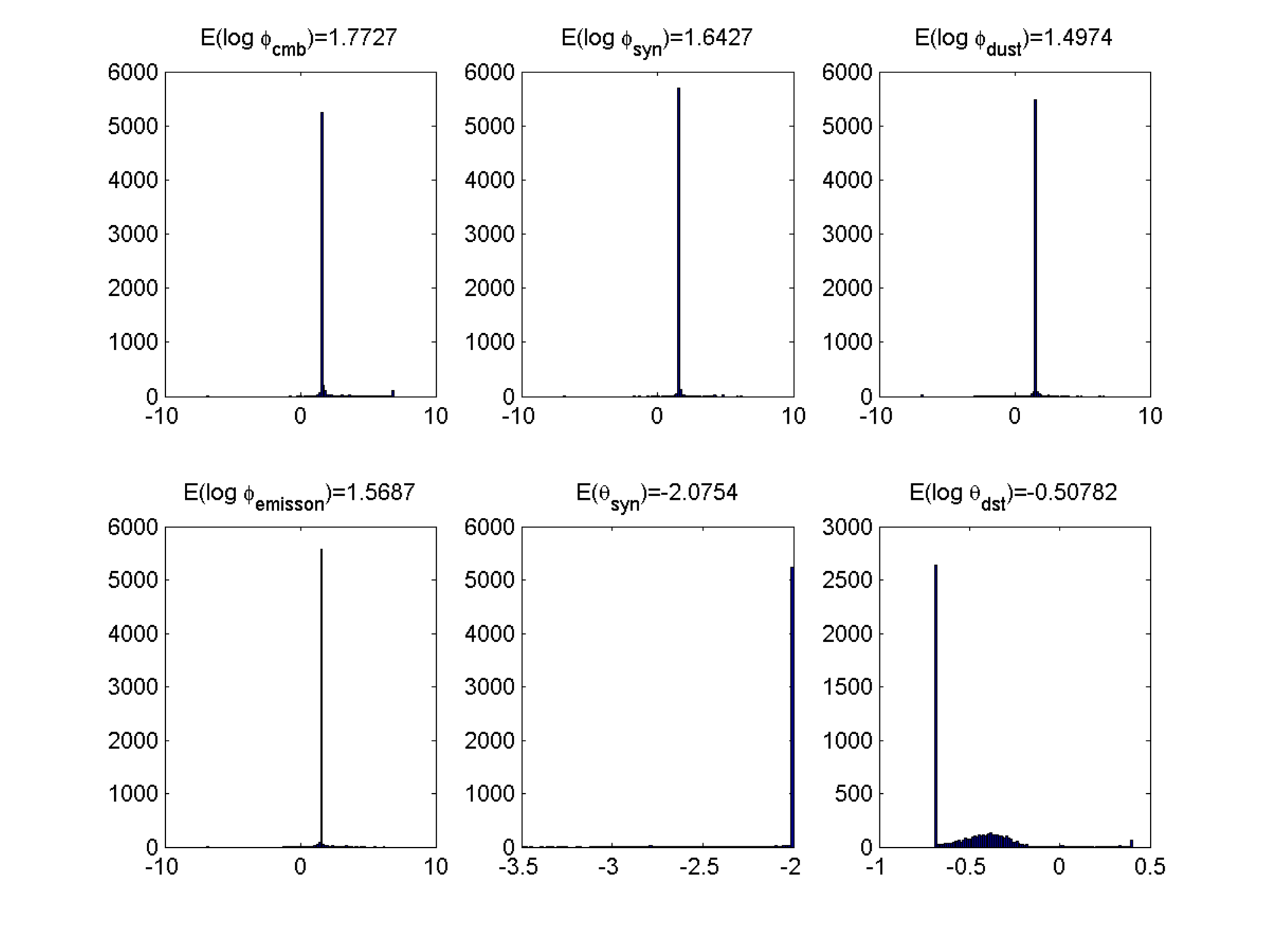}
\caption{\label{fig:params}Histograms of the posterior means of the elements of $\bm{\Psi}$, or their logarithm, over the 6144 blocks for the case where the prior mean of the $\phi_i$ is 10.   Average of the expectations are shown above each plot.}
\end{figure}

It has been noted that a successful separation can be achieved when the mean of the prior of the $\phi_i$ differs greatly but that these priors must have small variance, otherwise maps of the posterior expectations are not smooth and contain several large outlier pixel values.    Figure \ref{fig:cmb} shows the posterior expectation of CMB for 3 different priors on the $\phi_i$ with means of 1, 5 and 10, corresponding to weak, medium and strong spatial smoothness.  The effect of the prior on $\phi_i$ is clearly seen in the resulting separation.  Figure \ref{fig:other} shows the posterior means of the other separated sources where the prior mean of each $\phi_i$ is 10 (strong spatial smoothness), and Figure \ref{fig:params} shows histograms of the posterior means of the components of $\bm{\Psi}$, or their logarithm, over the 6144 blocks. Expectation of log parameter values are used in most cases for a clearer plot. It is seen that the prior on the $\phi_i$ has a strong influence on the posterior mean; the expectations of $\log(\phi_i)$ are remarkably consistent over the different blocks. 


\section{Discussion and Conclusion}
This paper has outlined a relatively straightforward method of approximating the posterior means of sources in a Gaussian source separation problem. By dividing the data into blocks, it can be used to conduct source separation in a reasonable time for data of the scale of WMAP and Planck. The blocking allows, particularly if parallel computation is available, an algorithm that is orders of magnitude faster than an MCMC approach.  

From the results in Figures \ref{fig:cmb} and \ref{fig:other}, the most obvious feature is that the galactic plane still causes considerable difficulties.  So while this is a completely automatic algorithm, it will still require manual processing about the galactic plane.  It is also noted that there does not appear to be an obvious block effect except near the galactic plane; a smooth reconstruction is in general obtained.  Block effects can be smoothed out in various ways, such as taking a moving average or averaging over overlapping blocks.

While the use of blocks is a necessary approximation, it is noted that a further restriction on the method is that the dimension of $\bm{\Psi}$ --- the vector of mixing matrix and prior source parameters --- must be small enough to allow a discrete grid to be stored and $q_W$ computed on it in a reasonable time.  It has been shown that this is tractable for 4 sources, with 6 parameters. The addition of an extra source adds 2 parameters to $\bm{\Psi}$ --- one for its mixing matrix column and one for the IGMRF prior --- so that the existing method becomes intractable for a much larger number of sources. For example, for 6 sources one would have 10 components in $\bm{\Psi}$, which would allow little more than a grid of 3 points along each dimension ($3^{10}$ points in ${\cal Q}$). For more sources, one option is to make a further approximation by forcing independence between the two set of components in $\bm{\Psi}$, the mixing matrix parameters $\theta_i$ and the IGMRF parameters $\phi_i$:
\[ p(\bm{\Psi} \, | \, \bm{Y}) \: = \: p(\bm{\theta} \, | \, \bm{Y}) \, p(\bm{\phi} \, | \, \bm{Y}), \]
and then compute independently $p(\bm{\theta} \, | \, \bm{Y})$ and $p(\bm{\phi} \, | \, \bm{Y})$ on separate grids of lower dimension. This would allow implementation of the algorithm to 6 sources comfortably.

Another observation is that the serial computation time is inversely related to the block size.  Suppose there are $K$ blocks.  The dominant computation is the Cholesky decomposition of the matrices $\bm{Q}^*_W(\bm{\Psi})$, which are of dimension $n_s J /K \times n_s J/K$. Computation of this decomposition is of order at worst $(n_s J /K)^3$, and at best $(n_s J / K)^2$ if  $\bm{Q}^*_W(\bm{\Psi})$ can be written as a band matrix, and there are $K$ of them, so in terms of $K$ the total computation time is of order $1/K$ to $1/K^2$.  So using smaller blocks is quicker, but this is clearly at the expense of an accurate approximation to $\mathbb{E}(\bm{S} \, | \, \bm{Y})$.  The block size of 512 pixels used here is a compromise with the longest computation time that we can justify.  For example, when using 24,576 blocks of 128 pixels, the computation time per block is about 0.75 seconds which gives a total serial computation time of about 5 hours. This compares with a total time of 72 hours for blocks of 512 pixels.

It is also noted that the identity used in Eq.\ \ref{eq:inla_basicapprox} can be used in cases where the likelihood $p(\bm{Y} \, | \, \bm{S}, \bm{\Psi})$ is not Gaussian.  In this case a Gaussian approximation to the denominator term $p(\bm{S} \, | \, \bm{Y}, \bm{\Psi})$ can be found by equating a mean and precision to its mode and curvature at the mode. The rest of the method of computing $\mathbb{E}(\bm{S}_W \, | \, \bm{Y}_W)$ is identical.  This is the integrated nested Laplace approximation of \cite{rue08} and has been shown to be very accurate in a wide range of latent Gaussian models. This would allow, for example, the use of non-linear relationships between $\bm{Y}$ and $\bm{S}$, or non-Gaussian measurement error.


\section*{Acknowledgement}
This work is supported by the STATICA project, funded by the Principal Investigator programme of Science Foundation Ireland, contract number 08/IN.1/I1879.


\bibliographystyle{chicago}
\bibliography{root}

\end{document}